\documentclass{JHEP3}

\usepackage{amsmath,amssymb,amsfonts}
\usepackage{graphicx}

\def\braket#1{\mathinner{\langle{#1}\rangle}}
\newcommand{\sbraket}[1]{\lbrack #1\rbrack}

\DeclareMathOperator{\tr}{tr}
\newcommand{\ii}{\mathrm{i}}
\newcommand{\Id}{\mathrm{I}}

\newcommand{\dalpha}{\dot{\alpha}}

\newcommand{\CP}{\mathbb{CP}}

\newcommand{\SO}{SO}

\title{Covariant representation theory of the Poincar\'e algebra and some of its extensions}
\author{Rutger Boels \\ Niels Bohr International Academy, Niels Bohr Institute\\ Blegdamsvej 17, DK-2100 Copenhagen, Denmark}

\preprint{\phantom{x}}
\keywords{Field Theories in Higher Dimensions, Extended Supersymmetry}

\abstract{There has been substantial calculational progress in the last few years for gauge theory amplitudes which involve massless four dimensional particles. One of the central ingredients in this has been the ability to keep precise track of the Poincar\'e algebra quantum numbers of the particles involved. Technically, this is most easily done using the well-known four dimensional spinor helicity method. In this article a natural generalization to all dimensions higher than four is obtained based on a covariant version of the representation theory of the Poincar\'e algebra. Covariant expressions for all possible polarization states, both bosonic and fermionic, are constructed. For the fermionic states the analysis leads directly to pure spinors. The natural extension to the representation theory of the on-shell supersymmetry algebra results in an elementary derivation of the supersymmetry Ward identities for scattering amplitudes with massless or massive legs in any integer dimension from four onwards. As a proof-of-concept application a higher dimensional analog of the vanishing helicity-equal amplitudes in four dimensions is presented in (super) Yang-Mills theory, Einstein (super-)gravity and superstring theory in a flat background.}

\begin{document}

\section{Introduction}
Amplitude calculations are the most direct channel to valuable information on the structure of any would-be theory of nature as they form the bridge between the formulation of the theory and experiment. A lesson learned in recent years is that amplitudes are much more constrained than previously appreciated by kinematic constraints and unitarity, especially if amplitudes are viewed as functions of complex momenta. These recent developments were initiated by Witten's twistor insights \cite{Witten:2003nn} for amplitudes in $\mathcal{N}=4$ super Yang-Mills theory at the origin of the moduli space. Much of the development since Witten's article especially on the analytic side has focused on amplitudes with four dimensional massless particles, with the proof of on-shell recursion relations for higher dimensional Yang-Mills theories \cite{ArkaniHamed:2008yf} and the proof of the BCJ \cite{Bern:2008qj} relations between amplitudes \cite{BjerrumBohr:2009rd, Stieberger:2009hq} as notable exceptions. That is not to say that more generic massive amplitudes are not interesting: in the weak sector everything including the vector bosons has a mass! The only known Lorentz-invariant unitary quantum field theories which contain massive four dimensional vector bosons are either spontaneously broken or, by dimensional decomposition, higher dimensional gauge theories. In this article we will focus on the latter leaving the former to \cite{companion}. Higher dimensional Yang-Mills theories also arise naturally within string theory and are used in high-loop calculations in $\mathcal{N}=4$ and $8$ for instance to keep track of all internal states in cuts \cite{Bern:2007dw}. A recently developed numerical approach to one loop QCD amplitudes also utilizes higher dimensional tree amplitudes \cite{Giele:2008ve}. Apart from these direct physical applications, from a more formal point of view it is an interesting question how much of recent progress really depends on being in four dimensions.

One central ingredient in all recent progress in analytic amplitude calculations has been an ability to keep precise track of the unique, discrete Lorentz quantum number for massless particles in four dimensions: the helicity. This is defined without reference to an explicit Lorentz frame, which is a necessary condition to be able to directly compare quantum numbers of different legs of a perturbative diagram. Helicity induces a natural ordering in complexity of scattering amplitudes of massless particles. The simplest ones with all helicities equal vanish for instance, as do the ones with one helicity opposite from the rest. This line of thought leads one to develop a covariant notation which keeps track of four dimensional helicity naturally: the spinor helicity method (see e.g. \cite{Dixon:1996wi} for a review and further references). The spinor helicity method leads amongst others to remarkably compact expressions for non-vanishing helicity amplitudes. Moreover, spinor helicity is the natural language to express the supersymmetry Ward identities for four dimensional massless particles. As such it is a central component of the recent progress in the calculation of field theory amplitudes. As will be worked out below, the spinor helicity method can be viewed as a simple consequence of \emph{covariant} representation theory of the Poincar\'e algebra and its supersymmetric extension. The main advantage of this point of view is that it is not restricted massless particles or to four dimensions and the main aim of this article is to develop this into a concrete tool. The relation to the usual lightcone-frame analysis will be indicated throughout. The methods developed here can be applied to scattering amplitudes in any higher dimensional (supersymmetric) field or string theory.

In this article we will focus first in section \ref{sec:Lorentzrepresentationsvectors} on the Poincar\'e quantum numbers for both massless and massive, single particle momentum eigenstates in higher dimensions. The basis of the analysis is a Lorentz-invariant definition of the little group generators which allows one to compare states on different legs of a diagram. This will be motivated from a quick look at four dimensional massive vector bosons. In the course of the construction the polarization vectors of both massive and massless vector states are obtained directly and by taking tensor products all representations of the Poincar\'e group in higher dimensions can be constructed. The analysis reduces to tracking the combined momentum and properly defined little group weight labels of the representation.  The extension to spinor representations is discussed next in section \ref{sec:purespinorhelicity}. Pure spinors arise naturally from this discussion and we provide a full dictionary to translate between vector and spinor representations. This amounts to a pure spinor helicity method based on tracking weight labels. The natural synthesis of both vector and spinor representations is the study of representation theory of the supersymmetry algebra in section \ref{sec:susy}. This leads directly to a method for obtaining Ward identities for scattering amplitudes in in principle any supersymmetric theory. Massive and massless particles are derived explicity. As a proof of concept application of the developed technology a certain class of vanishing amplitudes will be obtained in section \ref{sec:simpleapl}. A discussion and conclusions round of the presentation.

\section{Covariant Poincar\'e algebra representation theory}
\label{sec:Lorentzrepresentationsvectors}
\subsection{Warmup: the massive vector representation in four dimensions}
Massless vector representations of the Poincar\'e algebra in higher dimensions can always be decomposed into massive vector representations with respect to a chosen four dimensional sub-algebra. On a quest to find a generalization of the spinor helicity method to higher dimensions massive four dimensional vectors are therefore a good place to start and will serve as a cross-check. Furthermore, one of the applications of such a higher dimensional method would be an easy way to generate (BPS limits (\cite{Selivanov:1999ie}) of) amplitudes with massive vector bosons, which are of direct phenomenological importance. Spinor helicity methods have been discussed for massive four-dimensional states in several places in the literature (see e.g. \cite{Dittmaier:1998nn, Schwinn:2006ca} and references therein). Four dimensional `helicity' is for the theoretical purposes of this article not the right quantum number to study for massive particles as it is not a Lorentz invariant concept. The natural quantum number here is simply spin\footnote{More correct would be to use the term `massive spinor spin method' for the four dimensional contents of this subsection. In order to maintain backward compatibility this term will not be enforced here.}.

\subsection*{Explicit spin polarization vectors for massive vectors in four dimensions}
The massive spinor helicity formalism starts with the choice of a light-like vector $q$. For any massive vector there is a unique light-like vector $k^{\flat}$ such that
\begin{equation}\label{eq:decompmomentum}
k_{\mu} = k^{\flat}_{\mu} + \frac{k^2}{2 q\cdot k } q_{\mu}
\end{equation}
as long as
\begin{equation}
(q \cdot k) \neq 0 \ .
\end{equation}
This latter condition is important as it can lead to some complications when taking massless limits if not accounted for properly.

A natural choice of spin polarization axis if one wants to keep the link to the massless case as tight as possible is the spatial part of the lightcone vector $q$. The Lorentz-generator which implements rotations around this axis, denoted $R^1_q$, can be constructed from the Pauli-Lubanski vector in four dimensions,
\begin{equation}\label{eq:Lorentzrotprop}
R^1_q = \frac{q^{\mu} W_{\mu}}{2 q \cdot k} =  \frac{\epsilon_{\mu \nu \rho \sigma} q^{\mu} k^{\nu} \Sigma^{\rho \sigma} }{2 q \cdot k}
\end{equation}
That this is the correct generator can easily be checked by specializing to the rest-frame\footnote{In the vector representation $\left(\Sigma^{\rho \sigma}\right)_{\mu \nu} = i (\eta^{\rho}_{\, \mu} \eta^{\sigma}_{\, \nu} - \eta^{\sigma}_{\, \mu} \eta^{\rho}_{\, \nu})$. We choose a `mostly minus' metric convention.}. In the (smooth) massless limit this operator just measures the helicity. If desired, $R$ can be expressed in a more familiar form involving the space-like spin vector $s$ which can be isolated covariantly as
\begin{equation}
s_{\mu} = \frac{k_\mu}{m} - \frac{m q_{\mu}}{q \cdot k}  = \frac{k^{\flat}_{\mu}}{m} - \frac{m}{2 q\cdot k } q_{\mu}\ .
\end{equation}
This leads to the perhaps more familiar generator
\begin{equation}\label{eq:Lorentzrot}
R^1_q =  - \frac{1}{2} \frac{\epsilon_{\mu \nu \rho \sigma} s^{\mu} k^{\nu} \Sigma^{\rho \sigma} }{m}
\end{equation}
which however does not have a manifest massless limit. A third way to express the spin generator is to introduce a set of vectors which span $\mathbb{R}^{1,3}$,
\begin{equation}
q, \hat{q}, n_1, n_2
\end{equation}
where hat denotes parity conjugate. The vector $\hat{q}$ will not be needed in the following. The only non-zero inner products can be taken to be
\begin{equation}
q \cdot \hat{q} = n_1 \cdot n_1 = n_2 \cdot n_2 =1
\end{equation}
The rotation generator acts in the space orthogonal to both $k$ and $q$. A basis for this can be constructed out of $n_1$ and $n_2$,
\begin{equation}
\tilde{n}_i = n_i - q \frac{n_i \cdot k}{q \cdot k}
\end{equation}
and
\begin{equation}\label{eq:rotspecframe}
R^1_q = \tilde{n}_1^{\mu} \tilde{n}_2^{\nu} \Sigma_{\mu \nu}
\end{equation}
is obtained. Note that this generator commutes with $k$ by construction, as
\begin{equation}
\tilde{n}_i \cdot k = 0
\end{equation}
It can also be written as
\begin{equation}\label{eq:massivefourdgen}
R^1_q = \frac{1}{2} \frac{q^{\mu} n^{\rho}_1 n^{\sigma}_2 k_{[\mu} \Sigma_{\rho \sigma]}}{q \cdot k}
\end{equation}
where the brackets denote anti-symmetrization. This form will generalize most easily to higher dimensions.

The spin quantum number is determined from \eqref{eq:rotspecframe}, so the polarization vectors with eigenvalue $+$ and $-$ with respect to this rotation can be written as
\begin{equation}
e^+(k) = \frac{\tilde{n}_1 + \ii \tilde{n}_2}{\sqrt{2}} \quad \quad e^-(k) = \frac{\tilde{n}_1 - \ii \tilde{n}_2}{\sqrt{2}}
\end{equation}
This leaves the longitudinal polarization. By the usual massive vector field equation this must be orthogonal to $k$. It must also be orthogonal to the other two polarizations. Note that orthogonality to $k$ translates in a spontaneously broken gauge theory into unitary gauge. Completeness or an easy guess than lead to the spin vector $s$,
\begin{equation}
e_{\mu}^0(k) = s_{\mu}
\end{equation}
as the longitudinal polarization vector. The basis of states is normalized and the polarization sum can be performed explicitly,
\begin{equation}
e^+_{\mu} e^-_\nu + e^-_{\mu} e^+_\nu + e^0_{\mu} e^0_{\nu} = \eta_{\mu \nu} - \frac{k^\mu k^\nu}{m^2}
\end{equation}
which again shows clearly the unitary gauge.

Note that in the massless limit the longitudinal state is not well-normalized. This is connected to the fact that unitary theories of massive vector bosons only arise in conjunction with a (broken or unbroken) gauge symmetry. Typically, in a massless limit the gauge used to express the above polarization vectors becomes ill-defined and the singularity is therefore a gauge artefact. When embedded in a five dimensional gauge theory for instance, the above set of polarization vectors corresponds to a partial gauge choice for the fifth component of the gauge field,
\begin{equation}
A_5 = 0 ,
\end{equation}
which can obviously only be enforced as long as
\begin{equation}
k_5 \neq 0 \ .
\end{equation}
This momentum is equivalent to the mass from the four dimensional perspective as
\begin{equation}
k^2 = k_1^2 - k_2^2 - k_3^2 - k_4^2 - k_5^2 =0
\end{equation}
by the five dimensional mass-shell condition. Using the gauge freedom, other choices of polarization vectors are available which do have a smooth massless limit.

One of the advantages of a more covariant treatment of polarization is that it becomes possible to compare the polarization vectors of two different representations, i.e. two different particles. These obey in unitary gauge the simple but powerful equations
\begin{equation}\label{eq:simplebutpowerful}
e^+_{\mu}(i) e^{+,\mu}(j) = e^-_{\mu}(i) e^{-,\mu}(j) =0
\end{equation}
for polarization vectors of two different particles $i,j$ if the same polarization axis is chosen for these fields. The remaining inner products do not vanish. In the following we will see that this particular behavior determines several vanishing results for tree level scattering amplitudes. For longitudinal polarizations the story is slightly more complicated and will be spelled out in \cite{companion}.

\subsection*{Massive spinor helicity expressions in four dimensions}
For later use, let us briefly recall the massive spinor helicity expressions of the polarization vectors \cite{Dittmaier:1998nn}. Using massless spinor helicity methods, both lightcone vectors $q$ and $k^{\flat}$ in \eqref{eq:decompmomentum} can be written in terms of spinors $k^{\flat}_{\alpha}$, $k^{\flat}_{\dalpha}$, $q_{\alpha}$ and $q_{\dalpha}$ such that
\begin{equation}
k_{\alpha \dalpha} = k^{\flat}_{\alpha} k^{\flat}_{\dalpha} + \frac{k^2}{2 q\cdot k } q_{\alpha} q_{\dalpha}
\end{equation}
holds for the massive vector $k$. By the analogy with the massless case, the expressions for the polarization vectors in terms of massive spinor helicity variables can simply be guessed to be
\begin{equation}\label{eq:massivepolarization}
e^-_{\alpha \dalpha}(k) = - \ii \sqrt{2} \frac{q_{\alpha}k^{\flat}_{\dalpha}}{\braket{k q}}  \quad e^+_{\alpha \dalpha}(k) = - \ii \sqrt{2} \frac{q_{\dalpha}k^{\flat}_\alpha}{\sbraket{k q}} \quad e^0_{\alpha \dalpha} = \frac{k^{\flat}_{\alpha} k^{\flat}_{\dalpha}}{m} - \frac{m}{2 q\cdot k } q_{\alpha} q_{\dalpha}
\end{equation}
That these expressions transform as advertised under rotations can be checked explicitly in any frame but most straightforwardly in the rest frame.

\subsection{Massless vector representations in higher dimensions}
As alluded to above, the discussion of massive vectors in four dimensions is immediately applicable to five dimensional massless fields. Once this is known however, it turns out not to be too difficult to generalize to all higher dimensions. First even dimensions will be treated, leaving the extension to odd dimensions for later.

In general, on-shell massless vectors in $D$ dimensions have $D-2$ degrees of freedom, as one degree of freedom is removed by the field equation and another one by the gauge condition. The covariant representation theory of the Poincar\'e algebra becomes more complicated in higher dimensions for several reasons. A momentum eigenstate,
\begin{equation}
| k \rangle
\end{equation}
is the natural starting point, but there is no natural Pauli-Lubanski vector to fix the higher dimensional analogues of the spin quantum numbers. Instead, one obtains (see e.g. \cite{Pasqua:2004vq}) Pauli-Lubanski tensors,
\begin{equation}\label{eq:higherdimpaulilubanski}
W^{\mu \nu \rho}  =   k^{[\mu} \Sigma^{\nu \rho]}
\end{equation}
as objects which naturally commute with the momentum operator $k$. Here the brackets denote anti-symmetrization. The algebra of these tensors can be worked out in the lightcone frame and is that of the $ISO(D-2)$ group: the group of translations and rotations in $D-2$ Euclidean dimensions. Since there is no known physical interpretation of continuous spin quantum numbers in four dimensions (see e.g. \cite{Abbott:1976bb} and references therein), the eigenvalues of the translations will here also be set to zero. The remaining group is what is usually called the little group $SO(D-2)$. In four dimensions this group is Abelian and the canonical choice of generator then leads to the helicity quantum number. In higher dimensions the little group is non-Abelian and the representation theory of this little group will therefore entail a choice of a basis for the Cartan subalgebra. This is the major difference between four and higher dimensions. In fact, it is easy to recognize the spin generator of the massive four dimensional vector in equation \eqref{eq:massivefourdgen} as one of the Cartan generators of the massless higher dimensional vector. Note that this choice of Cartan generators is physical: in a four dimensional decomposition it corresponds to a choice of spin axis. In higher dimensions one way to think of it is that the choice of Cartan basis naturally corresponds to a series of polarization apparatuses in a higher dimensional experiment. Of course, there is a linear transformation from one basis to another.

The textbook analysis requires a choice of lightcone frame which can be avoided. As a natural extension of the analysis above choose an arbitrary orthonormal frame spanning $D$ even dimensions with signature $(1,D-1)$. Choose furthermore a real lightcone vector $q$ which has non-zero overlap with the real momentum $k$ and choose a lightcone gauge $q_{\mu} A^{\mu} = 0$. There is always a set of $D-2$ vectors $n_i$ such that
\begin{equation}
n_i \cdot q =0 =n_i \cdot \hat{q}
\end{equation}
and
\begin{equation}
n_i \cdot n_j = - \delta_{ij}.
\end{equation}
with hat denoting the parity conjugate. These can be modified such that they are orthonormal to the momentum $k$ as well as $q$,
\begin{equation}\label{eq:uncannyresemblancetoBCFW}
\tilde{n}^i  = n^i - q \frac{n^i \cdot k }{q \cdot k}
\end{equation}
In passing we note that this expression bears a resemblance to a BCFW-type shift \cite{Britto:2005fq}. Furthermore, we can choose a complex structure on the transverse directions to $q, \hat{q}$ spanned by the $n_i$ which induces a complex structure on the space orthogonal to $q$ and $k$ spanned by the $\tilde{n}_i$. To be explicit, take
\begin{align}
m^i & = \frac{\tilde{n}^{2i-1} + \ii \tilde{n}^{2i}}{\sqrt{2}} \\
\bar{m}^i & = \frac{\tilde{n}^{2i-1} - \ii \tilde{n}^{2i}}{\sqrt{2}}
\end{align}
with $i=1 \ldots \frac{\left(D-2\right)}{2}$.  We now have an orthonormal basis of $\mathbb{R}^{1,D-1}$ spanned by $D$ `light-like' vectors,
\begin{equation}\label{eq:completevectorbasis}
\{k,q,m^1, \bar{m}^1, \ldots, m^{\frac{\left(D-2\right)}{2}}, \bar{m}^{\frac{\left(D-2\right)}{2}} \}
\end{equation}
Just as in the massive vector case, it is easy to verify the $m_i$ and $\bar{m}_i$ have positive and negative eigenvalues respectively under
\begin{equation}
R^j_q = \ii m_j^{\mu} \bar{m}_j^{\nu} \Sigma_{\mu \nu} = \tilde{n}_{2j-1}^{\mu} \tilde{n}_{2j}^{\nu} \Sigma_{\mu \nu}
\end{equation}
and are orthogonal to both $q$ as well as $k$. Hence the vectors $m$ are the sought for on-shell polarization vectors, expressed in lightcone gauge with gauge vector $q$. Moreover, by construction the $R^j$ form a maximally commuting subset of the Poincar\'e generators. The relation to the Pauli-Lubanski tensors of equation \eqref{eq:higherdimpaulilubanski} is straightforward,
\begin{equation}
R^j_q = \frac{1}{2} \frac{q^{\mu}}{q \cdot k}  n_{2j-1}^\nu n_{2 j}^\rho   W_{\mu\nu\rho}
\end{equation}
These generators make explicit a choice of Cartan subalgebra generators which is Lorentz-invariant. We can label the on-shell states by the eigenvalues under the commuting system of conserved quantities $R^j_q$, i.e.
\begin{equation}
\left(R^j_q\right)^\mu_\nu e^\nu = h^j e^\mu
\end{equation}
Mathematically, these $h^j$ are the weight labels of the vector representation of the $\SO(D-2)$ group and we will use this term from here on. In the following the polarization vectors will be denoted as
\begin{equation}
e^{\mu}(k, h_1, \ldots h_{\frac{\left(D-2\right)}{2}}) \equiv  e^{\mu}(k, \vec{h})
\end{equation}
For the vector representation, the $h_i$ are either zero or $\pm 1$. Moreover, only one $h_i$ can be nonzero, so this leads to $D-2$ different polarization states, as expected.

The point of the above covariant construction is that now quantum numbers for two different states can be compared. For two different vector particles $i$ and $j$ in the same gauge defined by the vector $q$ the inner products between the polarization vectors are
\begin{equation}\label{eq:simplebutpowerfulII}
e(k_i, \vec{h}_i) \cdot e(k_j,  \vec{h}_j) = - \delta\left(\vec{h}_i +  \vec{h}_j\right) \ .
\end{equation}
This simple but powerful equation which is the generalization of \eqref{eq:simplebutpowerful} will cause the vanishing of several tree level amplitudes further on.

In the following all legs of an amplitude are assumed to share the same polarization complex structure, unless explicitly stated otherwise.

\subsection{Generalizations and remarks}

\subsubsection*{massive states}
The states above form a complete basis of vector states with specified eigenvalues under certain rotation operators. Hence, there is an immediate generalization to massive vector states. All polarization states are simply taken to be the same. The remaining polarization state needed follows from
\begin{equation}\label{eq:masslonghighdim}
\tilde{e}^{\textrm{UG}}_{\mu} = \frac{k_\mu}{m} - \frac{m q_{\mu}}{q \cdot k} \ .
\end{equation}
Again, the resulting basis is in unitary gauge.

\subsubsection*{odd dimensions}
The analysis in odd dimensions follows the same lines as for massive states in one dimension lower. In lightcone gauge the extra polarization state needed can be constructed using the space-like unit vector $n^D$ (taken to be orthogonal to all the vectors in one dimension lower) as
\begin{equation}
\tilde{e}^{\textrm{LG}}_{\mu}  =  n^{D}_{\mu} - q_{\mu}   \frac{n^{D} \cdot k}{q \cdot k}
\end{equation}
which is uncharged under any of the Cartan basis generators of the even-dimensional parent theory. The difference between this vector and the longitudinal component of the vector boson written above is a gauge transformation from the lightcone gauge used here to unitary gauge (most easily obtained by finding the gauge transformation for which $n_D \cdot \tilde{e}^{\textrm{LG}} =0$). As a case in point, one can consider the discussion above of massive four and massless five dimensional  vectors.

\subsubsection*{higher spins}
Although vector states where discussed above, it is clear that the discussion can be generalized immediately to general momentum eigenstate representations of the Poincar\'e group including (the generalization of) \eqref{eq:simplebutpowerfulII}. As an example, one can analyze massless spin-$2$ fields in any dimension which arise as the symmetric traceless part of the tensor product of on-shell spin-$1$ fields. In the $q$ lightcone gauge one obtains the $\left(\frac{(D-2)(D-1)}{2} -1 \right)$ different states as
\begin{equation}\label{eq:gravpolastates}
h^{\mu \nu}(k, \vec{h}_1 + \vec{h}_2) = \frac{1}{\sqrt{2}} \left(e^{\mu}(k, \vec{h}_1) e^{\nu}(k, \vec{h}_2) + e^{\nu}(k, \vec{h}_1) e^{\mu}(k, \vec{h}_2) \right)
\end{equation}
with the rule that the sum of labels should not vanish to enforce traceless-ness as follows from \eqref{eq:simplebutpowerfulII}. As implied by the vector indices, $h_1$ and $h_2$ are weight vectors of the vector representation of the little group. What is not immediately clear is how to treat spinor representations (of the covering Spin group), which is the subject of section \ref{sec:purespinorhelicity}.

It is actually not hard to find the non-Cartan generators of the little group in the notation defined above, as these are simply
\begin{equation}
e_{ij} \sim m_i \bar{m}_j \Sigma \ .
\end{equation}
with $i \neq j$. For completeness, the explicit generators of the translations of the translations of the $ISO(D-2)$ little group read in the language given above,
\begin{equation}
\frac{1}{2} \frac{\tilde{n}^i_{\mu} q_{\nu} k_{\rho} k^{[\mu} \Sigma^{\nu \rho]}}{q \cdot k} = \frac{1}{2}\tilde{n}^i_{\mu} k_{\nu}  \Sigma^{\mu \nu}
\end{equation}
Here the label runs from $1$ to $D-2$. As mentioned before, the eigenvalues of these operators are routinely set to vanish to avoid continuous spin quantum numbers. To make the representation theory more symmetric, the one continuous quantum number in the problem, $k$, can be treated as one of the two non-trivial eigenvectors of
\begin{equation}\label{eq:extrarotgen}
R^0_q =  \frac{q_{\mu} k_{\nu}   \Sigma^{\mu \nu}}{q \cdot k}
\end{equation}
the other being $q$. This generator will play a role in the analysis of the Spin group.

\subsubsection*{relation to light-cone frame}
The covariant analysis of the Poincar\'e quantum numbers presented above is of course closely related to the more familiar light-cone frame analysis of the on-shell degrees of freedom. To make this precise, note that we can always find a frame in which the light-like vectors $k$ and $q$ can be written as,
\begin{align}\label{eq:lightconeframevecs}
k = & \left( \begin{array}{cccc} 1 & 0 & \ldots & 1 \end{array} \right) k_0 \\
q = & \left( \begin{array}{cccc} 1 & 0 & \ldots & -1 \end{array} \right) q_0
\end{align}
For one light-like momentum (say $k$) this is trivial. The other (say $q$) is then obtained from utilizing both the rotations and translations of the $ISO(D-2)$ little group. The rotations give a generic form
\begin{equation}\label{eq:lightconeframevecsintermed}
q =  \left( \begin{array}{ccccc} a & b & 0 \ldots & c \end{array} \right)
\end{equation}
The finite Lorentz transformation from this form to the form of $q$ in \eqref{eq:lightconeframevecs} can be constructed from
\begin{equation}
\sim e^{\ii \epsilon \left(\Sigma_{01} +  \Sigma_{1 D}\right)}
\end{equation}
The generator leaves $k$ invariant and cubes to zero, which makes it straightforward to construct the explicit finite transformation. The $\tilde{n}$ vectors simply form a specific basis of the space transverse to both these vectors. They can therefore simply be assigned the numerical value
\begin{equation}
\left(n^i\right)_j = \delta^i_j
\end{equation}
in this particular frame. With these formulas in hand it is straightforward to evaluate this entire section in the lightcone frame.

\subsubsection*{gauge invariance and geometry}
Theories with higher dimensional massless vector bosons have a gauge invariance, i.e. amplitudes calculated using explicit polarization vectors and the LSZ procedure are invariant under
\begin{equation}\label{eq:changeofgaugepolvecs}
e^i_{\mu}(k) \rightarrow e^i_{\mu}(k) + g k_{\mu}
\end{equation}
with $g$ an arbitrary function. The above vectors are in the very natural lightcone gauge $q \cdot A =0$, so it is tempting to say that $q$ represents only a gauge choice. However, in the above $q$ is also part of the choice of Cartan generators or more abstractly a choice of complex structure: $q$ plays a dual role. Let one choice of $q$ be denoted $q_{\textrm{gauge}}$ and the other by $q_{\textrm{Cartan}}$. The discussion above is the choice $q_{\textrm{gauge}} = q_{\textrm{Cartan}}$. In four dimensions this distinction does not matter as the definition of helicity is independent of $q_{\textrm{Cartan}}$: the little group is Abelian.

Of course, $q_{\textrm{gauge}}$ can be changed independently of $q_{\textrm{Cartan}}$ by a gauge transformation. Since the $\vec{h}$ labels correspond to eigenvalues of generators of the little group (i.e. transformations which leave $k$ invariant) they are unchanged under the gauge transformations of equation \eqref{eq:changeofgaugepolvecs}. Changing both parameters $q_{\textrm{gauge}}$ and $q_{\textrm{Cartan}}$ simultaneously amounts to a combined gauge and non-Abelian little group transformation in higher dimensions. It would be very interesting to further explore this difference between four and higher dimensions.

One possible avenue of attack is the geometrical point of view. It is tempting to propose to replace the gauge invariance `choice of $q_{\textrm{Cartan}}$ in four dimensions' by equivalent choices of complex structure in higher dimensions. As a starting point, it can be noted that different choices of complex structure can be related by a complex coordinate transformation. Hence, the space of inequivalent structures on a Euclidean space is given by Lorentz transformations over complex transformations,
\begin{equation}
\sim \SO\left(D\right) / U\left(\frac{D}{2}\right)
\end{equation}
This is the space of pure spinors \cite{eliecartan}. As shown below, pure spinors will indeed arise naturally in the representation theory of the spin algebra. This does not involve the space of pure spinors directly though: this space should really be thought of as the higher dimensional analogue of twistor space. It would be very interesting to further elucidate the complex geometry hinted at by the above analysis.

\section{Covariant spin algebra representation theory}
\label{sec:purespinorhelicity}
The general discussion above will be extended to representations of the spin algebra in this section. Apart from a desire study covariant quantum numbers for spinors, this discussion leads directly to the sought-for higher dimensional spinor helicity method. Central is the realization that the polarization vectors and together the vectors $k$ and $q$ form a complete basis of the vector space. Using this fact yields through an analysis of the Clifford algebra a similar complete basis of spinors. The analysis yields furthermore a natural dictionary between vectors and spinors. The discussion has some features in common with a spinor helicity construction in six dimensions in \cite{Cheung:2009dc} which in fact inspired the present section. The exact relation to that work will be discussed below.

\subsection{Pure spinors and massless polarization vectors}
First a motivation for our spinor helicity construction is presented in a bottom up approach which will motivate the study of pure spinors. After this the complete setup is presented top down and the mathematically inclined reader may wish to skip to this.

\subsection*{Bottom up}
To first get an idea of what is required, note that given a solution to the massless Dirac equation for $k$ and one to its conjugate for $q$,
\begin{align}
\overline{\xi} q_{\mu} \gamma^{\mu}  & = 0 \\
k_{\mu} \gamma^{\mu} \psi  & = 0
\end{align}
it is easy to use the gamma matrix to construct vectors orthogonal to both $q_{\mu}$ and $k_{\mu}$ as
\begin{equation}\label{eq:probpolvec}
\sim \overline{\xi} \gamma^{\mu} \psi
\end{equation}
Note that it is clear that this is bound to be redundant in higher dimensions as the number of inequivalent solutions to the Dirac equation is half the minimal spinor dimension. In $10$ dimensions for instance, there are naively $64$ different states one can construct along the above lines. This degeneracy will be exposed in complete detail below. To do this, the weights of the vectors need to be obtained along the lines of the preceding section. Say the desired weight vector for the vector is $\vec{f}$. Given the complete, orthonormal basis in \eqref{eq:completevectorbasis}, one way to ensure this weight vector up to normalization is to demand that the above is orthogonal to all polarization states,
\begin{equation}
e_{\mu}(k, \vec{h}) \, \left( \bar{\xi} \gamma^{\mu} \psi \right) = 0
\end{equation}
for which the weight $h$ is not equal to the conjugate of the desired vector weight ($-\vec{f}$), and for that conjugate
\begin{equation}
e_{\mu}(k, -\vec{f}) \, \left( \bar{\xi} \gamma^{\mu} \psi \right) \neq 0
\end{equation}
should hold. A sufficient condition to satisfy these requirements is to have one of the following
\begin{align}\label{eq:choiceforspinors}
\overline{\xi} e_{\mu}(k, \vec{h}) \gamma^{\mu}  & = 0 \quad \quad \textrm{or}\\
e_{\mu}(k, \vec{h}) \gamma^{\mu} \psi  & = 0
\end{align}
for a specific weight $\vec{h}$. Now by the standard Clifford algebra a non-trivial state cannot be annihilated simultaneously by both a weight and its conjugate as
\begin{equation}
e_{\mu}(k, \vec{h}) e_{\nu}(k, -\vec{h}) \left( \gamma^{\mu} \gamma^{\nu} + \gamma^{\nu} \gamma^{\mu}\right) = \mathbb{I} \left(e_{\mu}(k, \vec{h}) e(k, -\vec{h})^{\mu} \right) =  \mathbb{I}
\end{equation}
as long as one of them is not $\vec{f}$. Hence, if
\begin{equation}
e_{\mu}(k, \vec{h}) \gamma^{\mu} \psi = 0
\end{equation}
then we must also have
\begin{equation}
\overline{\xi} e_{\mu}(k, -\vec{h}) \gamma^{\mu} = 0
\end{equation}
and vice-versa for the weights not equal to $\vec{f}$ or its conjugate. Furthermore,
\begin{align}
\overline{\xi} e_{\mu}(k, \vec{f}) \gamma^{\mu}  & = 0 \quad \quad \textrm{\underline{and}}\\
e_{\mu}(k, \vec{f}) \gamma^{\mu}\psi  & = 0
\end{align}
should hold. By the fact that we have a complete basis of the vector space, the resulting expression for the polarization vector,
 \begin{equation}
\sim \overline{\xi}(q) \gamma^{\mu} \psi(k)
\end{equation}
where the spinors obey the above constraints must be proportional to the sought-for vector state polarization vector with the weight labels given by ${\vec{f}}$. As noted before, there is a certain redundancy in the spinor helicity expressions: for a given weight of the vector representation, there are at least $\frac{D-2}{2}$ different combinations of operators which lead to the same quantum numbers. These follow from the different possible choices in \eqref{eq:choiceforspinors}.

Counting generators both spinor states $\overline{\xi}$ and $\psi$ are by construction annihilated by $\frac{D}{2}$ generators. Furthermore, the Lorentz vectors used for these $\frac{D}{2}$ generators form a set with vanishing inner products amongst themselves. Such a set is called isotropic (see e.g. \cite{Gualtieri:2003dx}), and in a flat $D$ real dimensional space-time with a metric, $\frac{D}{2}$ is the maximum number of them. Spinors which are annihilated by a maximally isotropic set of generators are called `pure' in the mathematical literature \cite{eliecartan}.

\subsection*{Top down}
The above bottom-up analysis and especially the annihilation conditions motivate a particular top-down approach. Mathematically the goal is to construct a representation of the Clifford algebra of $\mathbb{R}^{1, D-1}$,
\begin{equation}
\{\gamma^{\mu}, \gamma^{\nu} \} = 2 \eta^{\mu \nu}
\end{equation}
The standard way to do this can be found for instance in appendix B of \cite{Polchinski:1998rr} which will be modified here to suit the circumstances. Using the basis written in \eqref{eq:completevectorbasis} with the definitions
\begin{eqnarray}
\gamma^+_0  \equiv \frac{1}{2 q \cdot k} q_{\mu} \gamma^{\mu} & & \gamma^-_0  \equiv  k_{\mu} \gamma^{\mu}
\end{eqnarray}
and
\begin{eqnarray}\label{eq:completegammabasis}
\gamma^+_i  \equiv \frac{\ii}{\sqrt{2}} m^i_\mu \gamma^{\mu} & &  \gamma^-_i  \equiv \frac{\ii}{\sqrt{2}} \bar{m}^i_\mu \gamma^{\mu}
\end{eqnarray}
for $i=1,\ldots \frac{D-2}{2}$ this boils down to
\begin{equation}\label{eq:clifalgspecfram}
\{\gamma^{a}_i, \gamma^{b}_j \} =  \delta_{ij} \delta_{a, -b}
\end{equation}
where $a,b$ are $\pm$. The representation theory of this algebra is standard as it is $\frac{D}{2}$ copies of the fermionic harmonic oscillator. Note that in the basis \eqref{eq:completegammabasis} the Cartan generators in the spinor representation are simply
\begin{equation}
R^{i}_q = \frac{1}{2} [\gamma^+_i, \gamma^-_i] = \left(\gamma^+_i\gamma^-_i - \frac{1}{2}\right)
\end{equation}
for $i\geq 0$. The eigenvalues of $R^0_q$ determine if the spinor is a solution to the Dirac equation for either the vector $q_{\mu}$ or for $k_{\mu}$. The complete spinor representation is spanned by the tensor product of $\frac{D}{2}$ two dimensional spinor representations labeled by their eigenvalues under the $R^i_q$ generators for $i \geq 0$. The basis of spinors will be denoted by the set
\begin{equation}
\psi\left(\vec{h} \right)  \quad \quad  \xi\left(\vec{h} \right)
\end{equation}
here the components $h_i$ of the vector $\vec{h}$ are either plus or minus a half,
\begin{equation}
R_q^i \psi\left(\vec{h} \right) = h^i \psi\left( \vec{h} \right) \quad \quad  R_q^i \xi\left(\vec{h} \right) = h^i \xi\left( \vec{h} \right)
\end{equation}
with $i>0$. The difference between $\psi$ and $\xi$ is that
\begin{equation}
k_{\mu} \gamma^{\mu} \psi\left(\vec{h} \right) =0   \quad \quad q_{\mu} \gamma^{\mu} \xi\left(\vec{h} \right) = 0 \ .
\end{equation}

A pleasing feature of the construction is that all eigenvalues correspond directly to the $\frac{D}{2}$ equations
\begin{equation}\label{eq:spinorconditions}
\begin{array}{cl}\gamma_i^{2 h^i} \psi(\vec{h}) & = 0  \quad \quad \textrm{no sum}\ , \\
\gamma_{\mu} k^{\mu} \psi(\vec{h}) & = 0
\end{array}
\end{equation}
and
\begin{equation}\label{eq:spinorconditions2}
\begin{array}{cl}\gamma_i^{2 h^i} \xi( \vec{h}) & = 0  \quad \quad \textrm{no sum}  \\
\gamma_{\mu} q^{\mu} \xi(\vec{h}) & = 0 \ .
\end{array}
\end{equation}
Here notation has been abused slightly: $2h^i$ gives $\pm$ for the half-integer Dynkin weights of the spinor representation. A solution to these equations has the eigenvalues indicated by the weight labels. The spinors defined by \eqref{eq:spinorconditions} and \eqref{eq:spinorconditions2} are determined up to a scaling ambiguity for every solution,
\begin{equation}\label{eq:scalingambispinors}
\xi\left(\vec{h}\right) \rightarrow \beta_{\vec{h}} \xi\left(\vec{h}\right) \quad \psi\left(\vec{h}\right) \rightarrow \alpha_{\vec{h}} \psi\left(\vec{h}\right)
\end{equation}
Fixing this ambiguity amounts to choosing phase conventions.

The spinors constructed form a basis of all spinors in the dimension under study. As expected the full space of spinors contains a particular half subset which describes the solutions to the massless Dirac equation. As before, the conditions listed in \eqref{eq:spinorconditions} and \eqref{eq:spinorconditions2} form the definition of a pure spinor in $D$ dimensions: a pure spinor is annihilated by operators formed by contracting a maximally isotropic set of vectors with the gamma matrices. The setup is independent of the particular representation chosen for the gamma matrix algebra. The spinor representation can be reducible by a Weyl and/or a Majorana condition.

The chirality of the spinor is simply the product of the signs of $h^i$, taking into account $q$ and $k$ as well,
\begin{equation}\label{eq:chirinDdim}
\gamma_{D+1} = \ii \prod_{i=0}^{D-1} \gamma_i = -\ii \prod_{i=0}^{\frac{D}{2}} (\frac{2}{\ii} R^i_q)
\end{equation}
Note the lower bound on the product which takes into account the quantum number associated to rotations in the two-plane spanned by $q$ and $k$. A chirality constraint is easy to implement in a Lorentz invariant manner, so in applications it is usually advantageous to substitute chirality instead of one of the $R^i_q$ eigenvalues. Of course in four dimensions this directly leads to helicity. From the same observation it also follows that the helicity quantum number is independent of the choice of vector $q$: the same result holds for any $q$ (for which $q\cdot k\neq 0$).

Acting with a non-trivial generator $\gamma^{\pm}_i$ on a state will flip the quantum number. For example, acting with $\gamma_0^-$ on a $\xi(\vec{h})$ field yields a solution to the Dirac equation with weight vector $\vec{h}$,
\begin{equation}\label{eq:flippingktoq}
\gamma_0^-  \xi(\vec{h}) = k_{\mu} \gamma^\mu \xi(\vec{h}) \sim \psi(\vec{h})
\end{equation}
This follows from the fact that the operator commutes with the $R^i_q$ Cartan generators by construction and
\begin{equation}
k_{\nu} \gamma^{\nu}  k_{\mu} \gamma^{\mu}  = k_{\nu} k_{\mu} \{ \gamma^{\mu}, \gamma^{\nu} \} = k_{\nu} k^{\nu} \Id = 0
\end{equation}
on-shell.

Equation \eqref{eq:flippingktoq} can be made more precise in terms of the eigenspinors of the operators under study. To do this, the scalar spinor product will be needed which is defined by the usual combination of spinors involving $\gamma_0$,
\begin{equation}\label{eq:defspinprod}
\overline{\phi'} \phi \equiv \phi'^{\dagger} \gamma_0 \phi
\end{equation}
This spinor scalar product is inherited from Minkowski space reality conditions on the Lorentz transformations and the constraint that the product transforms as a scalar under these. The products between the eigenspinors can be evaluated in a specific frame (for instance in the tensor product representation) or more elegantly by using the algebra and the eigenspinor conditions. This shows the eigenspinors obey
\begin{equation}\label{eq:propfactspinprod}
\overline{\psi\left(h_0, \vec{h}\right)} \psi'\left(h'_0, \vec{h}'\right) \sim \, \delta\left(h_0 + h_0'\right) \delta\left(\vec{h} - \vec{h}'\right)
\end{equation}
The difference between $h_0$ and the weight vector $\vec{h}$ in the above formula is caused by the appearance of $\gamma_0$ in the scalar combination of spinors. For real momenta, we have
\begin{align}\label{eq:realmomconju}
\overline{\gamma_0^{\pm}} & =  \gamma_0^{\pm} \\
\overline{\gamma_i^{\pm}} & = - \gamma_i^{\mp}
\end{align}
It is important to realize that inner products for generic spinors in higher dimensions do not obey \eqref{eq:propfactspinprod}: this is a special feature of inner products with the $\xi$ spinors. These are special because the associated momentum $q$ is part of the definition of the Cartan generators.

With the scalar spinor product a representation of the generators \eqref{eq:completegammabasis} in terms of their eigenspinors can be written. In appendix \ref{app:eigenspinornorm} it is shown that spinors $\xi(\vec{h})$ and $\psi(\vec{h})$ can always be found such that
\begin{equation}\label{eq:sumsoverspinors}
\begin{array}{ccc}
\gamma^{\mu}  k_{\mu}  & = & \sum_{\vec{h}} \psi^{\vec{h}} \overline{\psi^{\vec{h}}} \\
\gamma^{\mu}  q_{\mu}  & = & \sum_{\vec{h}} \xi^{\vec{h}} \overline{\xi^{\vec{h}}} \\
\gamma^{\mu}  e_{\mu}\left(\vec{f}_i = \pm \delta_i^j\right) & = & \frac{\sqrt{2}}{\ii} \sum_{\vec{h} | h^j = \mp \frac{1}{2}} \left( \frac{\psi(\vec{h} + \vec{f}) \overline{\xi(\vec{h})}}{\overline{\xi(\vec{h})} \psi(\vec{h})} - \frac{\xi(\vec{h} + \vec{f}) \overline{\psi(\vec{h})}}{\overline{\psi(\vec{h} + \vec{f})} \xi(\vec{h} + \vec{f})} \right)
\end{array}
\end{equation}
These spinors obey the above spinor conditions \eqref{eq:spinorconditions2}, \eqref{eq:spinorconditions2}. The above formulas fix a phase convention, up to one phase for every $\xi$ and an simultaneous overall phase for all $\psi$. This phase convention will hold for all further formulas. In this convention
\begin{equation}\label{eq:niceeqfornormfacs}
\left(\overline{\psi(\vec{h})} \xi(\vec{h}) \right) \, \left( \overline{\xi(\vec{h})} \psi(\vec{h}) \right)= 2 q \cdot k
\end{equation}
holds, for every weight vector $\vec{h}$. For real momenta all spinor products are, up to the above remaining phase ambiguities, equal.

\subsubsection*{Vectors in terms of spinors}
Up to this point vectors contracted with the gamma matrices have been related to sums over spinors. In order to write polarization vectors directly in terms of the spinors the inverse to this is needed. In the `bottom-up' discussion above already a form for the polarization vectors was discussed. In the top down approach, the starting point is the realization that the gamma matrices are the Clebsch-Gordon coefficients for the tensor product of two spinor representations. This can easily be expressed as a relation between the vector and spinor representation generators,
\begin{equation}
\left(\Sigma_V^{\rho \sigma}\right)^\alpha_\beta \gamma^{\beta} = \Sigma_S^{\rho \sigma} \gamma^{\alpha} -  \gamma^{\alpha} \Sigma_S^{\rho \sigma}
\end{equation}
From this it follows that specifying the quantum numbers of the spinor states in \eqref{eq:probpolvec} specifies the quantum numbers of the corresponding polarization vector. As discussed before, this relation does not have a unique inverse. For every choice of polarization vector there are $\frac{D-2}{2}$ different ways of assigning quantum numbers to the spinor states if one restricts to the set of spinors defined above. This is straightforward expressed in the basis for the vectors \eqref{eq:completevectorbasis} and the corresponding basis of the gamma matrices, \eqref{eq:completegammabasis}. An overall constant can be fixed from \eqref{eq:simplebutpowerfulII}, using \eqref{eq:sumsoverspinors} to translate the vectors into spinor sums.

Concretely, this leads to the following dictionary between polarization vectors and spinors,
\begin{equation}\label{eq:purespinorpolari}
e^{\mu}(\vec{h}_1 - \vec{h}_2) = \frac{\ii}{\sqrt{2}} \frac{\overline{\xi(\vec{h}_1)} \gamma^{\mu} \psi(\vec{h}_2)}{\overline{\xi(\vec{h}_1)} \psi( \vec{h}_1)}
\end{equation}
Here $\vec{h}_1$ and $\vec{h}_2$ are weights of the spin representation of the little group, and their difference is restricted to be one of the weights of the vector representation. Contractions with one of the other vectors from the basis of the vector space are the correct ones by equation \eqref{eq:sumsoverspinors} and the properties of the spinor scalar product in equation \eqref{eq:propfactspinprod}.

There are similar expressions for $k_{\mu}$ and $q_{\mu}$,
\begin{align}
k_{\mu} & = \frac{1}{2} \overline{\psi(\vec{h})} \gamma_{\mu} \psi(\vec{h}) \\
q_{\mu} & = \frac{1}{2} \overline{\xi(\vec{h})} \gamma_{\mu} \xi(\vec{h})
\end{align}
for any weight vector $\vec{h}$.

The above derivation of equation \eqref{eq:purespinorpolari} strictly only holds for real momenta. In the complex case there can be a phase since \eqref{eq:simplebutpowerfulII} then does not fix vectors uniquely.

\subsection{Comparison to known spinor helicity results}
To elucidate the general structure found above let us work out two explicit examples in dimensions four and six and comment on the four dimensional limit of the higher dimensional cases.

\subsubsection*{Four dimensions}
In four dimensions vectors
\begin{equation}
q, n_1, n_2
\end{equation}
can always be found such that $q \cdot k \neq 0$. Following the discussion above, solutions to four different sets of equations are needed,
\begin{equation}
\begin{array}{l c r}
\left( \begin{array}{ccc} q_{\mu} \gamma^{\mu} & \xi\left( \frac{1}{2} \right) & = 0 \\
m^1_{\mu}          \gamma^{\mu} & \xi\left(\frac{1}{2} \right) & = 0 \end{array}   \right.   & \quad &
\left( \begin{array}{ccc} q_{\mu} \gamma^{\mu} & \xi\left(-\frac{1}{2} \right) & = 0 \\
\overline{m}^1_{\mu}   \gamma^{\mu} & \xi\left(-\frac{1}{2} \right) & = 0 \end{array}  \right. \\
\quad & \quad & \quad \\
\left( \begin{array}{ccc} k_{\mu} \gamma^{\mu} & \psi\left(\frac{1}{2} \right) & = 0 \\
m^1_{\mu}          \gamma^{\mu} & \psi\left(\frac{1}{2} \right) & = 0 \end{array}   \right.   & \quad &
\left( \begin{array}{ccc} k_{\mu} \gamma^{\mu} & \psi\left( -\frac{1}{2} \right) & = 0 \\
\overline{m}^1_{\mu}   \gamma^{\mu} & \psi\left(-\frac{1}{2} \right) & = 0 \end{array}  \right.
\end{array}
\end{equation}
By the analysis above, the half integer quantum number indicates the eigenvalue under $R^1_q$. Since $\xi$ and $\psi$ have a definite eigenvalue under $R^0_q$, this generator gives equivalent information as the four dimensional chirality operator in $4$ dimensions by equation \eqref{eq:chirinDdim}. Hence a definite eigenvalue under $R^1_q$ translates immediately into the requirement that the spinor is chiral. This  is another manifestation of the Abelian nature of the little group. Vice-versa, fixing a chirality fixes the eigenvalue under $R^1_q$: the helicity eigenvalue for a pure spinor.

In terms of more conventional spinor helicity notation it is easy to write down all the different sought-for pure spinors,
\begin{eqnarray}\label{eq:basisfourdspinors}
\begin{array}{c}
\xi\left(-\frac{1}{2} \right)  = \left(\begin{array}{c} 0  \\ q^{\alpha} \end{array}\right) \\
\xi\left(\frac{1}{2} \right)  = \left(\begin{array}{c} q_{\dalpha} \\ 0 \end{array}\right) \end{array} & &
\begin{array}{c}
\psi\left(\frac{1}{2} \right)   = \left(\begin{array}{c} 0 \\ k^{\alpha} \end{array}\right) \\
\psi\left(-\frac{1}{2} \right)  = \left(\begin{array}{c} k_{\dalpha} \\ 0 \end{array}\right)
\end{array}
\end{eqnarray}
It will be checked below that this is the right normalization. For these the first two lines of \eqref{eq:sumsoverspinors} follow easily
\begin{equation}
k_{\mu} \gamma^{\mu} = \left(\begin{array}{cc} 0 & k_{\mu} \sigma^{\mu}_{\dalpha \alpha} \\ k_{\mu} \overline{\sigma}^{\mu, \alpha \dalpha} & 0 \end{array} \right) = \left(\begin{array}{cc} 0 &  k_{\dalpha} k_\alpha \\ k^\alpha k^{\dalpha} & 0 \end{array} \right)
\end{equation}
as is standard practice. Note that from
\begin{equation}
\sigma^{\mu} k_{\mu} = \left(\begin{array}{cc} k_0 + k_3 & k_1 + \ii k_2 \\  k_1 - \ii k_2 & k_0 - k_3 \end{array}\right)
\end{equation}
the parity conjugated expression can be derived by a similarity transform as
\begin{equation}
k_{\mu} \overline{\sigma}^{\mu} = \left(\begin{array}{cc} 0 & 1 \\  -1 & 0 \end{array}\right) \left(\begin{array}{cc} k_0 + k_3 & k_1 + \ii k_2 \\  k_1 - \ii k_2 & k_0 - k_3 \end{array}\right)^T \left(\begin{array}{cc} 0 & -1 \\  1 & 0 \end{array}\right) \ .
\end{equation}
This is independent of reality conditions on the momenta.

The inner product on the spinor space gives as non-vanishing inner products
\begin{eqnarray}
\overline{\xi\left(-\frac{1}{2} \right)} \psi\left(-\frac{1}{2} \right)  = q^{\dalpha} k_{\dalpha} & \quad &
\overline{\psi\left(\frac{1}{2} \right)} \xi\left(\frac{1}{2} \right)  =  k^{\dalpha} q_{\dalpha} \\
\overline{\psi\left(-\frac{1}{2} \right)} \xi\left(-\frac{1}{2} \right)  =  k_{\alpha} q^{\alpha} & \quad  &
\overline{\xi\left(\frac{1}{2} \right)} \psi\left(\frac{1}{2} \right)  =  q_{\alpha} k^{\alpha}
\end{eqnarray}
With the spinors in hand, the usual spinor helicity polarization vectors for the vectors follow, e.g.
\begin{align}
e^+_{\mu} \gamma^{\mu} & = \frac{\sqrt{2}}{\ii} \frac{\psi\left(k , \frac{1}{2} \right) \overline{\xi\left(q , -\frac{1}{2} \right)}}{\overline{\xi\left(q , -\frac{1}{2} \right)} \psi\left(k , -\frac{1}{2} \right)} - \frac{\sqrt{2}}{\ii} \frac{\xi\left(q , \frac{1}{2} \right) \overline{\psi\left(k , -\frac{1}{2} \right)} }{\overline{\psi\left(k , \frac{1}{2} \right)} \xi\left(q ,  \frac{1}{2} \right)} \\
&  = - \ii \left(\begin{array}{cc}  0 & \sqrt{2} \frac{k_{\dalpha} q_{\alpha}}{\braket{k q}}  \\ \sqrt{2} \frac{q^{\alpha}k^{\dalpha}}{\braket{k q}} & 0 \end{array} \right)
\end{align}
For every polarization of the gluon there is just one independent expression in terms of the spinors, as the chiral and anti-chiral expressions are related by parity. In other words, there is a consistent truncation to chiral or anti-chiral spinors.

\subsubsection*{Six dimensions}
Since five dimensions was already discussed above by studying massive spinor helicity for four dimensional vectors, the next example of higher dimensional spinor helicity can be found in six dimensions. An interesting side-product of the analysis will be an explicit solution of the four dimensional Dirac equation with prescribed spin eigenvalues through four dimensional massive spinor helicity methods. In this subsection we adhere to the phase conventions set in ordinary spinor helicity.

A natural set of gamma matrices geared toward the study of a decomposition w.r.t. the first four dimensions is
\begin{equation}
\gamma_{\mu} =  \left(\begin{array}{cc} 0 & \gamma_\mu \\  \gamma_\mu & 0 \end{array} \right)
\end{equation}
for $\mu = 1,\ldots, 4$ and $5$ with the usual four dimensional gamma matrices taken for convenience in the chiral representation and
\begin{equation}
\gamma^6 = \left(\begin{array}{cc} 0 & \Id_{4x4} \\ -\Id_{4x4} & 0 \end{array} \right)
\end{equation}
for the sixth gamma matrix. Consider the Dirac equation in six dimensions,
\begin{equation}
\gamma^{\mu} k_{\mu} \psi_6 = 0
\end{equation}
where $\psi_6$ is a complex chiral spinor,
\begin{equation}
\psi_6 = \frac{\left(1+ \gamma_7\right)}{2} \psi_6
\end{equation}
with $8$ real components which, in the representation of the gamma matrices chosen here, can be expressed in terms of a `four dimensional' complex spinor $\psi_4$ as,
\begin{equation}
\psi_6 =  \left( \begin{array}{c} \psi_4 \\ \vec{0}\end{array} \right)
\end{equation}
where $\vec{0}$ is a four dimensional zero vector. In the four dimensional language the non-trivial part of the Dirac equation reads
\begin{equation}
\left(\begin{array}{cc} 0 & k_{\mu} \sigma^\mu \\ k_{\mu} \bar{\sigma}^\mu & 0 \end{array} \right) \psi_4 + \left(\begin{array}{cc}  (k_6 + \ii k_5) \textrm{Id}_{2x2} & 0 \\ 0 &  (k_6 - \ii k_5 ) \textrm{Id}_{2x2}\end{array} \right) \psi_4 = 0
\end{equation}
which is the (complex) massive Dirac equation in four dimensions. The complex spinor can be decomposed in terms of two Weyl spinors as
\begin{equation}
\psi_4 = \left(\begin{array}{c} \phi_{\dalpha} \\ \phi^{\alpha} \end{array} \right)
\end{equation}
This leads to
\begin{equation}
k_{ \dalpha \alpha} \phi^{\alpha} = \left( k^{\flat}_{\dalpha} k^{\flat}_{\alpha} + \frac{k_4^2}{2 q\cdot k }  q_{\dalpha} q_{\alpha}\right) \phi^{\alpha} = (k_6 + \ii k_5) \phi_{\dalpha}
\end{equation}
and
\begin{equation}
k^{\alpha \dalpha} \phi_{\dalpha}  =  \left(k^{\flat, \alpha} k^{\flat, \dalpha} + \frac{k_4^2}{2 q\cdot k } q^{\alpha} q^{\dalpha}\right) \phi_{\dalpha} = (k_6 - \ii k_5) \phi^{\alpha}
\end{equation}
The field equations imply of course the higher dimensional massless Klein-Gordon equation for both of these components,
\begin{equation}
\left(\left(\sum_{\mu=1}^4 k_{\mu} k^{\mu}\right)  - k_5^2 - k_6^2\right) \phi_{\alpha} = \left(\left(\sum_{\mu=1}^4 k_{\mu} k^{\mu}\right) - k_5^2 - k_6^2\right) \phi^{\dalpha} = 0
\end{equation}
where $\left(\sum_{\mu=1}^4 k_{\mu} k^{\mu}\right)$ is of course the four dimensional mass.

In the previous section a complete basis of four dimensional spinors was obtained in \eqref{eq:basisfourdspinors} spanned by $q$ and $k^{\flat}$, and here it is natural to expand everything in sight in them. It turns out to be easiest to define new Lorentz invariant fields $\phi^{\dot{1}}, \phi^{\dot{2}}, \phi_1$ and $\phi_2$ as,
\begin{align}\label{eq:covariantizaitonof6dspinor}
\phi^{\alpha} & \equiv (k_6 - \ii k_5) \frac{q^{\alpha}}{\braket{q k^{\flat}}} \phi_1 + k^{\flat, \alpha} \phi_2 \\
\phi_{\dalpha} & \equiv (k_6 + \ii k_5) \frac{q_{\dalpha}}{\sbraket{q k^{\flat}}} \phi^{\dot{1}} + k^{\flat}_{\dalpha} \psi^{\dot{2}}
\end{align}
From this definition the resulting system of equations is easily solved as
\begin{equation}\label{eq:sol6dmassless4dmassive}
\phi_1 = \phi^{\dot{2}} \quad \quad \phi_2 = \phi^{\dot{1}}
\end{equation}
resulting into a solution for the six-dimensional massless Dirac equation,
\begin{equation}\label{eq:solsixDdiraceq}
\psi_4 = \phi_1   \left(\begin{array}{c} (k_6 - \ii k_5) \frac{q_{\alpha}}{\braket{q k^{\flat}}}  \\ k^{\flat, \dalpha} \end{array} \right) + \phi_2 \left(\begin{array}{c}  k^{\flat}_{\alpha}   \\  (k_6 + \ii k_5) \frac{q^{\dalpha}}{\sbraket{q k^{\flat}}}  \end{array} \right)
\end{equation}
This solution has $4$ real degrees of freedom from the two complex numbers $\phi_1$ and $\phi_2$. Having solved the six dimensional chiral Dirac equation, let us turn to the quantum numbers. There are two in principle, but one can be determined in terms of the other by the six dimensional chirality condition of equation \eqref{eq:chirinDdim}. Hence, determining the four dimensional spin of the particle using $R^{1}_q$ determines the quantum number of $R^2_q$. This quantum number corresponds in the four dimensional limit to rotations in the $5,6$ plane. In the (four dimensional) rest frame it is easy to verify that the spin up and down solutions simply correspond to $\phi_1=0$ or $\phi_2=0$ respectively. Concretely,
\begin{align}
\psi\left(\frac{1}{2}, \frac{1}{2}\right) & = \left(\begin{array}{c} k^{\flat}_{\dalpha} \\ (k_6 - \ii k_5) \frac{q^{\alpha}}{\braket{q k^{\flat}}}  \\ \vec{0} \end{array} \right)\\
\psi\left(-\frac{1}{2}, -\frac{1}{2}\right) & = \left(\begin{array}{c}   (k_6 + \ii k_5) \frac{q_{\dalpha}}{\sbraket{q k^{\flat}}} \\ k^{\flat,\alpha}   \\ \vec{0} \end{array} \right)
\end{align}
where a normalization has been chosen to satisfy,
\begin{equation}
k_{\mu} \gamma^{\mu} \frac{1}{2} (1+\gamma_{7})= \psi\left(\frac{1}{2}, \frac{1}{2}\right) \overline{\psi\left(\frac{1}{2}, \frac{1}{2}\right)} + \psi\left(-\frac{1}{2}, -\frac{1}{2}\right) \overline{\psi\left(-\frac{1}{2}, -\frac{1}{2}\right)}
\end{equation}
Note that the solutions written in \eqref{eq:sol6dmassless4dmassive} are Lorentz invariant as these coefficients are Lorentz invariant.

In addition to the solution written above there are anti-chiral solutions to the six dimensional Dirac equation (with weight vectors $\left(\frac{1}{2}, -\frac{1}{2}\right)$ and $\left(-\frac{1}{2}, \frac{1}{2}\right)$). The above expressions are the $\psi$ type pure spinors, while the $\xi$ type follows from the four dimensional analysis by the choice of aligning $q$ with the four dimensional space: one lifts these spinors to either a chiral or anti-chiral six dimensional spinor, i.e.
\begin{eqnarray}
\begin{array}{cc} \xi\left(\frac{1}{2}, \frac{1}{2}\right) & = \left(\begin{array}{c} \left(\vec{0}\right) \\ \left(\begin{array}{c} q_{\dalpha} \\ 0 \\0 \end{array} \right)\end{array} \right)\\ \phantom{p} & \phantom{p} \\
\xi\left(-\frac{1}{2}, -\frac{1}{2}\right) & =  \left(\begin{array}{c} \left(\vec{0}\right) \\ \left(\begin{array}{c}  0 \\0 \\ q^\alpha\end{array} \right)\end{array} \right) \end{array} & & \begin{array}{cc}
\xi\left(\frac{1}{2}, -\frac{1}{2}\right) & = \left(\begin{array}{c}  \left(\begin{array}{c} q_{\dalpha} \\ 0 \\0 \end{array} \right) \\ \left(\vec{0}\right) \end{array} \right)\\ \phantom{p} & \phantom{p} \\
\xi\left(-\frac{1}{2}, \frac{1}{2}\right) & =  \left(\begin{array}{c} \left(\begin{array}{c} 0 \\0 \\ q^\alpha \end{array} \right) \\  \left(\vec{0}\right)\end{array} \right) \end{array}
\end{eqnarray}
Note the inversion of chirality for a given weight vector with respect to the $\psi$ spinors.

When constructing polarization vectors a choice has to be made which chirality of the spinors is used. This leads to two expressions in terms of spinors for every polarization vector. Concretely,
\begin{align}
e^{\mu}\left(k,(1,0)\right) & = - \ii \sqrt{2} \frac{\overline{\xi\left(\frac{1}{2}, \frac{1}{2}\right) }\gamma^{\mu} \psi\left(-\frac{1}{2}, \frac{1}{2}\right)}{\overline{\xi\left(\frac{1}{2}, \frac{1}{2}\right)} \psi\left(\frac{1}{2}, \frac{1}{2}\right)}   \\
e^{\mu}\left(k,(-1,0)\right)  & = - \ii \sqrt{2} \frac{\overline{\xi\left(-\frac{1}{2}, -\frac{1}{2}\right) }\gamma^{\mu} \psi\left(\frac{1}{2}, -\frac{1}{2}\right)}{\overline{\xi\left(\frac{1}{2}, \frac{1}{2}\right)} \psi\left(\frac{1}{2}, \frac{1}{2}\right)}   \\
e^{\mu}\left(k,(0,1)\right)  & = - \ii \sqrt{2} \frac{\overline{\xi\left(\frac{1}{2}, \frac{1}{2}\right) }\gamma^{\mu} \psi\left(\frac{1}{2}, -\frac{1}{2}\right)}{\overline{\xi\left(\frac{1}{2}, \frac{1}{2}\right)} \psi\left(\frac{1}{2}, \frac{1}{2}\right)}  \\
e^{\mu}\left(k,(0,-1)\right) & = - \ii \sqrt{2} \frac{\overline{\xi\left(-\frac{1}{2}, -\frac{1}{2}\right) }\gamma^{\mu} \psi\left(-\frac{1}{2}, \frac{1}{2}\right)}{\overline{\xi\left(\frac{1}{2}, \frac{1}{2}\right)} \psi\left(\frac{1}{2}, \frac{1}{2}\right)}
\end{align}
These follow simply from specifying the general expression written in \eqref{eq:purespinorpolari}.

Reverting these expressions to four dimensional spinor helicity notation gives for both choices of chirality
\begin{align}
e_{\mu}\left(k,(1,0)\right)  & = - \ii \sqrt{2} \frac{q_{\dalpha}k^{\flat}_\alpha}{\sbraket{k^{\flat} q}} \sigma^{\alpha \dalpha}_{\mu} \\
e_{\mu}\left(k,(-1,0)\right) & = - \ii \sqrt{2} \frac{q_{\alpha}k^{\flat}_{\dalpha}}{\braket{k^{\flat} q}} \sigma^{\alpha \dalpha}_{\mu} \\
e^{\mu}\left(k,(0,1)\right)  & = - \ii /{\sqrt{2}} \left( (\ii k_5 + k_6) \frac{q^{\mu}}{q \cdot k} + \ii n^{\mu}_5 + n^{\mu}_6 \right)\\
e^{\mu}\left(k,(0,-1)\right) & = - \ii /{\sqrt{2}} \left( (\ii k_5 - k_6) \frac{q^{\mu}}{q \cdot k} - \ii n^{\mu}_5 + n^{\mu}_6 \right)
\end{align}
Here the convention $\sigma_{\mu} =0$ has been employed for $\mu=5,6$. These expressions show clearly what these polarization vectors are, as the first two reproduce the massive vector spinor helicity expressions written in equation \eqref{eq:massivepolarization}. The other two are clearly the generalization of what are the scalar states in a four dimensional reduction.

As referred to earlier recently a spinor helicity method was proposed in six dimensions in \cite{Cheung:2009dc}. The main difference of the above approach to that paper is that there the little group generators were left implicitly defined. An advantage of this is that one obtains amplitudes with all possible polarizations at the same time. It does go against intuition built in four dimensions that some amplitudes are more simple than others when the Poincar\'e quantum numbers are taken into account. Moreover, for concrete applications to polarized massive scattering amplitudes in four dimensions the four dimensional spin needs to be specified as the choice of spin axis is physical.

\subsubsection*{Four dimensional limit from higher dimensions}
One important check of all results obtained using pure spinor helicity is to evaluate the calculation with the restriction that all momenta lie in a common four dimensional plane (which includes $q$). In this limit, the calculation should reproduce known four dimensional answers. Let the four dimensions be spanned by $k,q,n_1$ and $n_2$. The pure spinor constraints which do not involve $\gamma_1^{\pm}$ or $k_{\mu} \gamma^{\mu}$ are all independent of the momenta. In this case one can solve the pure spinor constraints with a simple natural Ansatz,
\begin{equation}\label{eq:fourdimlimit}
\phi = \phi_4\left(h_0, h_1 \right) u\left(h_2 \ldots h_{\frac{D-2}{2}}\right)
\end{equation}
where $\psi_4$ is a four dimensional pure spinor as analyzed above. The function $u$ is momentum independent and transforms in the spinor representation of $\SO(D-4)$ with the indicated weights. For generic momenta, this Ansatz of course never solves the conditions.

\subsection{Remarks, generalizations and applications}
Some readers might be familiar with pure spinors from a string theory context where the space of pure spinors forms an essential ingredient of Berkovits' approach to the covariant quantization of the superstring \cite{Berkovits:2000fe}. In that part of the string literature pure spinors are defined by
\begin{equation}
0= \psi \gamma^\mu  \psi  = \overline{\psi} \gamma_0 \gamma^\mu  \psi\quad \forall \mu \ .
\end{equation}
The relation of this condition to the above should be clear as using any of the matrices from the complete basis in \eqref{eq:completegammabasis} will always flip one and only one quantum number.

\subsubsection*{massive}
With the technology developed, it is also possible to write down spinor helicity-type expressions for the polarization vectors for massive particles in dimensions higher than four. The explicit expression for the polarization vectors are the same as in the massless case, with $k$ replaced by $k^{\flat}$ in the pure spinors. This forms a basis of polarization vectors of a massive vector boson in higher dimensions. Furthermore, the remaining longitudinal polarization follows from the spinor version of \eqref{eq:masslonghighdim}. As before, the resulting expression are in unitary gauge. Massive solutions to Dirac's equation are discussed below.

\subsubsection*{obtaining explicit spinors in a fixed frame}
For a numerical implementation or explicit analytic calculations in a special frame, it is important to obtain easily evaluated expressions for all the spinors in the above analysis. Fortunately, there is a straightforward calculational path as the above analysis reduces analyzing the spinors to representations of the fermionic harmonic oscillator. As noted above, a special feature of any harmonic oscillator is that given one state in the theory, all others may be constructed by acting with the generators which do not annihilate the given state. Let this state be given by
\begin{eqnarray}\label{eq:setofhighestweighteqsspinors}
q_{\mu} \gamma^{\mu} \xi(\vec{h}_0) = 0 & \quad \quad & \gamma_i^{+} \xi(\vec{h}_0)  = 0
\end{eqnarray}
for $i=1,\ldots \frac{D-2}{2}$. These equations correspond to the Dynkin weights $\vec{h}_0=\left( \frac{1}{2},\ldots,\frac{1}{2}\right)$. By the first equation, the subsequent equations are independent of the momenta,
\begin{equation}
\left(\frac{n_{\mu}^{2i-1} - \ii n_{\mu}^{2i}}{\sqrt{2}}\right) \gamma^{\mu} \xi =0
\end{equation}
For a numerical implementation a frame has to be chosen. A natural choice of frame is given by
\begin{eqnarray}\label{eq:specframeexpl}
q = \frac{1}{\sqrt{2}} \left(1,0,\ldots,-1\right) & \quad & n^{j}_{\mu}  = \delta^j_{\mu}
\end{eqnarray}
For a given representation of the gamma matrices, a straightforward set of equations given by evaluating \eqref{eq:setofhighestweighteqsspinors} in this frame has to be solved. Moreover, in a gamma matrix representation in which $\gamma_{D+1}$ is diagonal, it is easy to obtain chiral fermions. This makes one of the $\frac{D}{2}$ constraints superfluous by equation \eqref{eq:chirinDdim}. Solving the equations gives one explicit solution to the set of conditions in \eqref{eq:spinorconditions2}. The other solutions are now easily obtained by acting with $\gamma^-_i$ on this solution for $i=1,\ldots \frac{D-2}{2}$. The weight vectors are easily read of, as these operators flip the corresponding $h_i$ quantum number.

Up to now nothing has depended on the momentum of the Dirac spinor we would like to obtain, and all calculations can be performed in an initialization phase. In particular a residual phase ambiguity of the $\xi$ spinors is fixed in the initialization phase. Solutions to the Dirac equation follow by application of the slashed momentum, as
\begin{equation}
k_{\mu} \gamma^{\mu} \xi(q, \vec{h}) = b \psi(k,\vec{h})
\end{equation}
In the next step a normalization has to be worked out for the spinors to correspond to the spinors in the text. The numerical price of constructing these spinors is therefore fairly low.

\subsubsection*{lightcone frame}
Of course, the above construction method is closely related to a lightcone frame analysis. Equivalence is achieved by specifying the momentum also in lightcone form. There is a slight philosophical issue here with the order in which choices are made: if an arbitrary coordinate system is chosen first and represented as in equation \eqref{eq:specframeexpl}, the inner products of the momentum $k$ with these vectors are Lorentz-invariants and generically non-zero. Hence it is impossible to specify the momentum as
\begin{equation}
k = \left(1,0,\ldots,-1\right)
\end{equation}
in general. The solution to this problem is to chose a different frame: one in which the momentum dependent vectors $\tilde{n}$ obey
\begin{equation}
\tilde{n}^{j}_{\mu}  = \delta^j_{\mu}.
\end{equation}
This choice is however not universal as the $\tilde{n}$ vectors depend on the momentum. With this choice the lightcone frame analysis is equivalent to the one above.

\subsubsection*{explicit pure spinor solutions to the massive Dirac equation}
\label{sec:Lorentzrepresentationsmasses}
The method used to solve the six-dimensional Dirac equation above also yields a solution to the four dimensional massive Dirac equation with prescribed quantum numbers. This side-product can be generalized to higher dimensions using the pure spinors defined above. Consider the massive Dirac equation
\begin{equation}\label{eq:higdimdirsolvmass}
\left(k_{\mu} \gamma^\mu + m \right) \phi=0 \ ,
\end{equation}
Using equations \eqref{eq:decompmomentum} and \eqref{eq:sumsoverspinors}  $k_{\mu} \gamma^\mu$ can be written as
\begin{equation}\label{eq:higdimmassmomexp}
k_{\mu} \gamma^\mu = \sum_{\vec{h}} \left(\psi\left(k^{\flat}, \vec{h} \right) \overline{\psi\left(k^{\flat}, \vec{h}\right)} + \frac{m^2}{2 q \cdot k} \xi\left(\vec{h} \right) \overline{\xi\left(\vec{h}\right)} \right)
\end{equation}
In principle everything can again be expanded in terms of the pure spinor basis and coefficients can be read off. However, it is fairly easy to guess that for prescribed quantum numbers $\vec{h}$ one should simply take an Ansatz of the form
\begin{equation}\label{eq:ansatmassdirhighdim}
\phi(k,\vec{h}) = x_1 \psi\left(k^{\flat},\vec{h}\right) + x_2 \xi\left(\vec{h}\right)
\end{equation}
Note that these two spinors have opposite chirality. Inserting this into \eqref{eq:higdimdirsolvmass} and utilizing \eqref{eq:higdimmassmomexp} yields two equations for two variables
\begin{eqnarray}
x_1 m  + x_2 \overline{\psi(\vec{h})} \xi(\vec{h})    & = 0 \\
x_1 \frac{m^2}{2 q \cdot k}  \overline{\xi(\vec{h})} \psi(\vec{h}) + x_2 m  & = 0
\end{eqnarray}
However, this system is degenerate because of equation \eqref{eq:niceeqfornormfacs} and hence the system has a non-trivial solution. With this solution \eqref{eq:ansatmassdirhighdim} constitutes a solution to the Dirac equation with prescribed quantum numbers $\vec{h}$. Note that there are $2^\frac{D-2}{2}$ complex solutions of this type, as expected.

\subsubsection*{some notational simplifications for multiple particles}
For multiple particles it is advantageous to use a notation
\begin{equation}
\psi\left(k_1,\vec{h}\right) = 1^{\vec{h}}
\end{equation}
and
\begin{equation}
\xi\left(q,\vec{h}\right) = q^{\vec{h}}
\end{equation}
The conjugate spinors are denoted
\begin{equation}
\overline{\psi\left(k_1,\vec{h}\right)} = \overline{1^{\vec{h}}}
\end{equation}
and
\begin{equation}
\overline{\xi\left(\vec{h}\right)} = \overline{q^{\vec{h}}}
\end{equation}
In addition, the set of weights with definite chirality for the $\psi$ spinors will be denoted $K^{\pm}$,
\begin{equation}
\frac{1}{2}\left(1 \mp \gamma_D \right) \psi\left(k,\vec{h}\right) = 0 \quad \quad \forall \vec{h} \in K^{\pm}
\end{equation}
The $\xi$ spinors have opposite chiralities for these weights,
\begin{equation}
\frac{1}{2}\left(1 \pm \gamma_D \right) \xi\left(\vec{h}\right) = 0 \quad \quad \forall \vec{h} \in K^{\pm}
\end{equation}
In the following spinors and conjugate spinors will be treated as independent, corresponding to complex momenta.

\section{Covariant supersymmetry algebra representation theory}\label{sec:susy}
The two sections above are covariant versions of the representation theory of the Poincar\'e and the spin algebra. It is known that (under certain restrictions) the only non-trivial extensions of these symmetry algebras which admit non-trivial scattering behavior are either supersymmetric or conformal, and this motivates an extension of the discussion to these cases. In this article we focus on the former since momentum eigenstates are not well-behaved representations of the conformal algebra\footnote{These states are not normalizable over the conformal compactification of spacetime, $S^4$. This fact leads to subtle violations of conformal symmetry in the scattering amplitudes \cite{Mason:2009sa, ArkaniHamed:2009si}.}. In addition, one of the foreseen applications of a higher dimensional spinor helicity method is the study of high loop maximally supersymmetric amplitudes in four dimensions through unitarity-based techniques and these higher dimensional theories are not conformal. Higher dimensional scattering amplitudes with Poincar\'e supersymmetry are of course a very natural subject from the superstring point of view as well. The main reason that an extension of the above analysis for free fields to supersymmetric theories is very interesting is that full control over the on-shell fields leads immediately to on-shell Ward identities for the scattering amplitudes.

\subsection{Supersymmetry Ward identities}
There are in principle two different methods to derive supersymmetry Ward identities for scattering amplitudes in four dimensions in the literature. The first \cite{Grisaru:1976vm} proceeds by writing down an explicit Lagrangian field theory formulation and finding the supersymmetry variations subject to LSZ truncation. This can be applied in principle to any field theory for which an off-shell formulation is known and has been applied to the  $4D$  fundamental massive multiplet in \cite{Schwinn:2006ca}. This derivation method is redundant however, as it requires off-shell information for the calculation to proceed while the S-matrix elements are purely on-shell. To our knowledge, the Ward identity in \cite{Schwinn:2006ca} is actually the \emph{only} representation for which these identities are known explicitly besides the $4D$ massless ones. A variant of this derivation method proceeds through superstring theory in a flat background \cite{Stieberger:2007jv}, which although fully on-shell has its own idiosyncrasies.

A second derivation \cite{Grisaru:1977px} only uses the on-shell supersymmetry algebra and special properties of massless spinors in four dimensions to derive the identities. Schematically, if the vacuum of a theory is supersymmetric under a generic supersymmetry variation
\begin{equation}
Q|0\rangle =0
\end{equation}
and the S-matrix commutes with $Q$, then the correlation function from which the amplitude is calculated obeys
\begin{equation}\label{eq:derivWardids}
0 = \langle 0 | S \, \Phi_{\textrm{in}} \, Q \,| 0 \rangle = \langle 0  | S \, [Q, \Phi_{\textrm{in}}] | 0 \rangle =  \langle 0 | S \, Q \,| \textrm{in} \rangle
\end{equation}
Here $\Phi_{\textrm{in}}$ on the vacuum prepares a specific fermionic, free in-state which can be chosen as desired. Since this state is free it is given by the tensor product of different creation operators. The commutator with $Q$ of the in-state is therefore a sum over commutators of the creation operators. However,
\begin{equation}
[Q,c(k,\vec{h})] |0\rangle = Q |k,\vec{h} \rangle
\end{equation}
Therefore, if the action of $Q$ on the \emph{free and on-shell} state $|k,\vec{h} \rangle$ is known then an identity between scattering amplitudes with different field content arises. This action should follow from the on-shell supersymmetry algebra, but the usual representation theory is only worked out in the lightcone frame of one particle. As before the covariant action of $Q$ on the single particle states is needed.

In a more modern version (see e.g. \cite{Dixon:1996wi}), the above derivation can be made precise efficiently through four dimensional spinor helicity methods. The latter approach will be generalized below to all representations of the supersymmetry algebra without central charges in any dimension using the pure spinors defined above. The derivation is based on a simple observation which has been made above: one can use pure spinor helicity methods to reduce the analysis in a canonical way to a rest-frame or lightcone frame problem. Representation theory in this frame is standard in principle, although the treatment of weight vectors presented below in this context appears to be new as far as the author is aware.

Consider the supersymmetry algebra in $D$ dimensions,
\begin{equation}\label{eq:susyinanyd}
\{Q, \overline{Q}\} = 2 k_{\mu} \gamma^{\mu}
\end{equation}
and first restrict to massless particles. For every momentum eigenstate there is a basis of pure spinors. Hence $Q$, $\overline{Q}$ and $k_{\mu} \gamma^{\mu}$ can be expressed in terms of them. For the momentum this is \eqref{eq:sumsoverspinors}, while for the supersymmetry generators we define,
\begin{equation}\label{eq:generatorexpansion}
Q \equiv \sqrt{2} \sum_{\vec{h}} \left(Q_{\vec{h}} k^{\vec{h}} + Q'_{\vec{h}} q^{\vec{h}} \right)
\end{equation}
and its conjugate,
\begin{equation}\label{eq:generatorexpansionII}
\overline{Q} \equiv \sqrt{2} \sum_{\vec{h}} \left(\overline{Q}_{\vec{h}} \overline{k^{\vec{h}}} + \overline{Q}'_{\vec{h}} \overline{q^{\vec{h}}} \right)
\end{equation}
For now the sum runs over all fermionic weights. The $Q_h$ generators can be extracted from this by way of equation \eqref{eq:propfactspinprod},
\begin{equation}
Q_{\vec{h}} = \frac{\overline{q^{\vec{h}}} Q}{\sqrt{2} \overline{q^{\vec{h}}} k^{\vec{h}}}
\end{equation}
and
\begin{equation}
Q'_{\vec{h}} = \frac{\overline{k^{\vec{h}}} Q}{\sqrt{2} \overline{k^{\vec{h}}} q^{\vec{h}}}
\end{equation}
Formulae for the conjugate are similar. Comparing left and right of the equation in \ref{eq:susyinanyd} gives as a non-trivial part of the algebra
\begin{equation}\label{eq:susyfermharmosc}
\{ Q_{\vec{h}}, \overline{Q_{\vec{h}}} \} = 1
\end{equation}
which can be recognized as several copies of the fermionic harmonic oscillator algebra, whose representation theory is standard. The trivial part of the algebra determines
\begin{equation}\label{eq:trivialrepsmassless}
Q'_{\vec{h}} = 0 \quad \quad \overline{Q'_{\vec{h}}} = 0
\end{equation}
Counting non-trivial $Q$-generators gives half of the minimal spinor dimension.

Since the operators $Q_h$ and $Q'_{h}$ can be expressed in terms of Lorentz invariant spinor products, their action can be studied in any frame, and in particular in the lightcone frame. The action of the general operator $Q$ then follows from equation \eqref{eq:generatorexpansion}. In other words, this equation can be used to generalize standard representation theory to any frame. This is the key observation of this section. As in the textbook approach, one should take care that the highest weight state forms a complete representation of the Poincar\'e group, which in general can be reducible by Weyl or Majorana conditions.

The actions of $Q_{\vec{h}}$ and $\overline{Q_{\vec{h}}}$ on a state change the quantum numbers in predictable ways,
\begin{equation}
Q_{\vec{h}} |k,\vec{g}\rangle = |k,\vec{g}+\vec{h}\rangle
\end{equation}
and
\begin{equation}
\overline{Q_{\vec{h}}} |k,\vec{g}\rangle  =  |k,\vec{g}-\vec{h}\rangle
\end{equation}
It is straightforward to check that in four dimensions the analysis reproduces the well-known transformations of the massless multiplet.

In the argument of this section there is no limitation on the dimensionality of space from a mathematical perspective. Of course, the spinor representations in space-times above eleven dimensions will contain spin $>2$ fields in the four dimensional limit which are notoriously hard to incorporate in a physical interacting quantum field theory with a finite number of fields. This leads to the usual physical constraint to consider a maximum of $10$ or $11$ dimensions.

Massive representations follow along similar lines, after using \eqref{eq:decompmomentum} to express the massive momentum in two massless ones. Note that the resulting massive representation contains twice as many states as the massless ones, since in this case \eqref{eq:trivialrepsmassless} does not hold any more. The representation theory of this in four dimensions will be further explored in \cite{companion}. Here also equivalence to the results of \cite{Schwinn:2006ca} will be shown. Multiple supersymmetries without central charges are easily incorporated into the argument. Central charges make the analysis technically more complicated, but are not a fundamental obstruction: higher dimensional massless representations can always be decomposed \cite{Fayet:1978ig} into lower dimensional BPS representations. A simple check of this is simply counting the states.

\subsubsection*{$\mathcal{N}=1$ vector multiplet in $D=6$}
\begin{figure}[h]
  \begin{center}
  \includegraphics[height=0.6\textwidth]{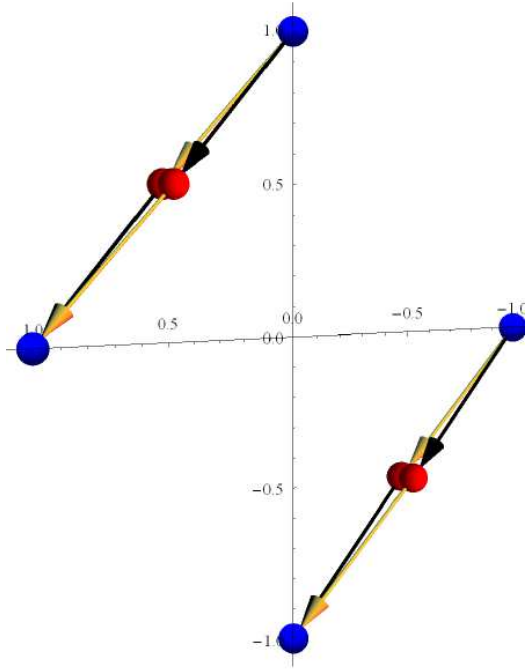}
  \caption{the $\mathcal{N}=1$, $D=6$ vector multiplet in weight space with a degeneracy lifted for illustration purposes}
  \label{fig:d6n1multiplet}
  \end{center}
\end{figure}

As a first example, consider the massless vector multiplet for $\mathcal{N}=1$ in $D=6$. The minimal spinor in this dimension is a complex Weyl spinor with $4$ complex, $8$ real degrees of freedom. On-shell, there are only four real degrees of freedom as demonstrated explicitly in section \ref{sec:purespinorhelicity}. Equivalently, there are only two (complex) non-trivial supersymmetry generators and their conjugates,
\begin{eqnarray}
Q_{\frac{1}{2}, \frac{1}{2}} & & \overline{Q_{\frac{1}{2}, \frac{1}{2}}}\\
Q_{-\frac{1}{2}, -\frac{1}{2}} & & \overline{Q_{-\frac{1}{2}, -\frac{1}{2}}}
\end{eqnarray}
Without loss of generality, here a choice of chiral over anti-chiral spinors was made. The representation theory proceeds by picking a highest weight state, which can be degenerate. For the vector representation one can pick the vector states  $|k,\vec{h}\rangle=|1,0\rangle$ and $|k,\vec{h}\rangle= |0,1\rangle$ for instance and immediately write down the multiplet structure in weight space. This is illustrated in the figure \ref{fig:d6n1multiplet}.

To interpret the figure it is important to realize that the spinors are generically complex, so the state $|\frac{1}{2}, -\frac{1}{2}\rangle$ actually has two real degrees of freedom: it is degenerate. It is instructive to compare this degeneracy to the same phenomenon in the explicit solutions to the six dimensional Dirac equation found above in equation \eqref{eq:solsixDdiraceq}. If the momentum is restricted to a four dimensional plane, it is clear that this reproduces $\mathcal{N}=2$, $D=4$ with the phase supplying the R-symmetry quantum number.

\subsection*{Coherent state representation}
The above analysis can be used to obtain relations between explicit amplitudes, following \cite{Grisaru:1977px} directly. For the purposes of the present paper there is a more beautiful argument which proceeds through coherent states, inspired by the analysis of \cite{ArkaniHamed:2008gz}. It amounts to the observation that when all supersymmetry transformations are applied to a chosen highest weight state, the resulting expression is a sum over all states in the representation. The possibility of writing coherent states follows directly from the fermionic harmonic oscillator algebra \eqref{eq:susyfermharmosc}.

The only subtle point in the analysis is the choice of highest weight state. Choose one, say
\begin{equation}
|k, \vec{t} \rangle
\end{equation}
This state is annihilated by half of the supersymmetry algebra. Let $L_{\vec{t}}$ denote the set of the weights of the $Q$ generators which annihilate the chosen highest weight state,
\begin{equation}
Q_{\vec{h}} |k, \vec{t} \rangle =0 \,\, \forall \vec{h} \in L_{\vec{t}}
\end{equation}
The complement of this set within the set of (chiral) weights will be denoted by $\overline{L}_{\vec{t}}$. For these
\begin{equation}
\overline{Q_{\vec{h}}} |k, \vec{t} \rangle =0 \,\, \forall \vec{h} \in \overline{L}_{\vec{t}}
\end{equation}
holds, as any state in the multiplet is always annihilated by half of the generators. For every weight introduce two series of fermionic coherent state parameters,
\begin{equation}
\eta_{\vec{h}} \quad \overline{\eta_{\vec{h}}}
\end{equation}
For every choice of highest weight state there is a set of coherent state parameters of half the minimal spinor dimension,
\begin{equation}
\{\eta_t(\vec{h}) \} = \left( \begin{array}{cc} \eta_{\vec{h}} & \quad \textrm{if} \quad \vec{h} \in  \overline{L}_{\vec{t}} \\ \overline{\eta_{\vec{h}}} & \quad \textrm{if} \quad \vec{h} \in  L_{\vec{t}} \end{array} \right.
\end{equation}

A coherent state can now be defined as
\begin{equation} \label{eq:cohpurespin}
|k \{\eta_t(\vec{h}) \} \rangle = e^{\sum_{\vec{l} \in \overline{L}_{\vec{t}}} Q_{\vec{l}} \eta_{\vec{l}} + \sum_{\vec{l} \in L_{\vec{t}}} \overline{\eta}_{\vec{l}} \overline{Q_{\vec{l}}} } |k, \vec{t} \rangle
\end{equation}
The non-trivial terms in the terminating power series with parameters  $\eta_{\vec{l}}$ and $\overline{\eta}_{\vec{l}}$ contain precisely the states of the full multiplet. This is straightforward to verify in the lightcone frame and since the generators are defined in a Lorentz invariant way, the result holds in every frame. As usual, care has to be taken that the supersymmetry multiplet forms a complete representation of the Poincar\'e group. In the six dimensional case for instance, a coherent state representation of $\mathcal{N}=1$ would consist of a sum over two coherent states defined above. Therefore, the above is most useful in the case the supersymmetry multiplet coincides with a representation of the Poincar\'e group. Examples of this include $\mathcal{N}=1$ in $D=10$ for the Yang-Mills multiplet and $\mathcal{N}=1$ in $D=11$ for the gravity multiplet or any of their dimensional reductions.

The supersymmetry transformations act naturally on the states in \eqref{eq:cohpurespin}. A generic susy transformation on the massless state can be written as
\begin{equation}
\mathfrak{Q} = \exp{\left(\overline{Q}\chi + \overline{\chi} Q \right)} = \exp{\left(\sqrt{2} \sum_{\vec{h}} \left(\overline{k^{\vec{h}}} \chi \right) \, \overline{Q_{\vec{h}}} + Q_{\vec{h}} \, \left( \overline{\chi} k^{\vec{h}} \right) \right)}
\end{equation}
In this expression the vanishing terms from \eqref{eq:trivialrepsmassless} have been put to zero. This transformation acts on the coherent state as
\begin{multline}\label{eq:cohsusytrafo}
\mathfrak{Q} |k \{\eta_t(\vec{h}) \} \rangle  = e^{\sqrt{2} \sum_{\vec{l} \in L_{\vec{t}}}  \left( \overline{\chi} k^{\vec{h}} \right) \eta_{\vec{l}}  + \sqrt{2} \sum_{\vec{l} \in \overline{L}_{\vec{t}}} \left(\overline{k^{\vec{h}}} \chi \right) \overline{\eta}_{\vec{l}} } \\
 e^{\sum_{\vec{l} \in \overline{L}_{\vec{t}}} Q_{\vec{l}} \left(\eta_{\vec{l}} + \sqrt{2} \left( \overline{\chi} k^{\vec{h}} \right) \right) + \sum_{\vec{l} \in L_{\vec{t}}} \left(\overline{\eta}_{\vec{l}} + \sqrt{2} \left(\overline{k^{\vec{h}}} \chi \right) \right) \overline{Q_{\vec{l}}} } |k, \vec{t} \rangle
\end{multline}
In other words, half of the transformation shifts the coherent state variables, while the other half leads to a phase shift.

The coordinates $k, \eta, \overline{\eta}$ in the coherent states are coordinates of a direct generalization of Nair's on-shell superspace \cite{Nair:1988bq} to all dimensions. Nair's space is very closely connected to the super-twistor space $\CP^{3|4}$, so the pure spinor expression \eqref{eq:cohpurespin} should at least in spirit be very closely related to the pure spinor superspaces studied in the literature.

\subsection{Example: $\mathcal{N}=1$ SYM in $D=10$}
\begin{figure}[t]
  \begin{center}
  \includegraphics[width=0.5\textwidth]{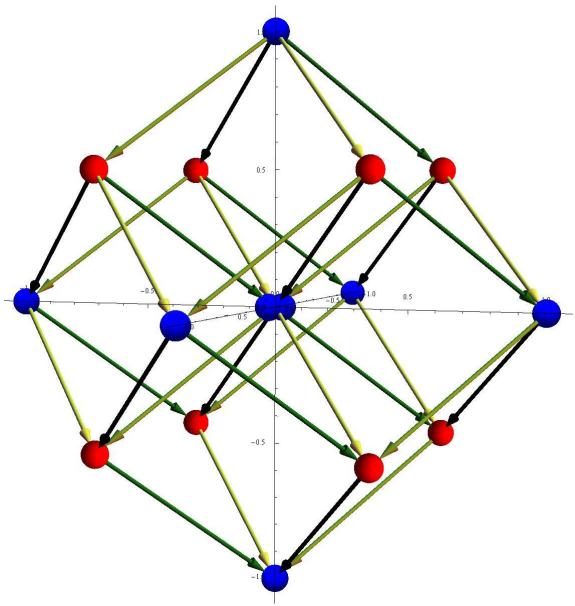}
  \caption{the $\mathcal{N}=1$, $D=10$ vector multiplet in (projected) weight space}
  \label{fig:n4multiplet}
  \end{center}
\end{figure}

As a cross check and example of the transformations, let us study $\mathcal{N}=1$ super Yang-Mills theory in $D=10$. The multiplet structure in weight space of the fundamental massless multiplet in this theory is illustrated in figure \ref{fig:n4multiplet}. The minimal spinor obeys both a Weyl as well as a Majorana condition. Although the weight vectors contain four entries in this dimension, due to the Weyl chirality constraint specifying three for the fermions suffices. Hence the only ambiguous point in the graphical representation is the origin, which is doubly degenerate in this three dimensional picture. For illustration purposes, in the picture this degeneracy has been lifted by hand. The arrows indicate the supersymmetry transformations needed to obtain the other states in the theory from the highest weight state at $(0,0,1)$. Different colors represent different transformations.

The coherent states discussed above can of course be considered in $10$ dimensional SYM. In the four dimensional limit \eqref{eq:cohpurespin} the ten-dimensional expressions reduces to the coherent states of \cite{ArkaniHamed:2008gz} directly, as can be seen from \eqref{eq:fourdimlimit}. Actually, one gets $6$+$8$ additional possible coherent state representations not studied there by taking one of the scalar states or the fermions as the highest weight which involve a mixture of $Q$ and $\overline{Q}$ operators. The main difference in higher dimensions is the role played by the $\xi$-type spinors in the definition of the correct operators: these do not represent a gauge degree of freedom here. The consequences of this for actual amplitudes are the subject of the next section.

\subsubsection*{Field theory derivation}
Although the above derivation method of the supersymmetry transformations is completely general since it only depends on algebra representation theory, one might still be interested in what the field theory equivalent derivation would look like. This can be worked out in the instructive example of the fundamental representation of $\mathcal{N}=1$, $D=10$. The Lagrangian of the $10$ dimensional super Yang-Mills theory is given by
\begin{equation}
\mathcal{L} = \tr \int d^{10}\!x \frac{1}{4} F_{\mu\nu} F^{\mu\nu} + \frac{1}{2} \overline{\phi} \gamma_{\mu} \left(\partial^{\mu} + g A^{\mu}\right)\phi
\end{equation}
where $\phi$ is an anti-chiral $10$ dimensional Majorana spinor which transforms in the adjoint of the gauge group. Based on symmetries, the form of the rigid supersymmetry transformations can easily be guessed to be
\begin{align}
\delta_{\chi} A^{\mu} & =  \overline{\chi} \gamma^{\mu} \phi \\
\delta_{\chi} \phi & =  F^{\mu\nu} \Sigma_{\mu \nu} \chi
\end{align}
where $\chi$ is here a chiral Majorana spinor which parameterizes the supersymmetry transformation,
\begin{equation}
\mathfrak{Q} = \overline{\chi} Q
\end{equation}
Up to gauge transformations and equations of motion these variations generate the supersymmetry algebra. Since we will only be interested in on-shell quantities, this is not a restriction as the extra terms can simply be dropped. Explicit creation operators can be written down including quantum numbers for the different components of the in state, and their supersymmetric transformation properties can be studied. Explicitly, these read
\begin{equation}\label{eq:extvecstate10D}
g^{\vec{h}}(k) \equiv e^{\mu}(k, \vec{h}) A_{\mu}
\end{equation}
for the vector and
\begin{equation}\label{eq:extfermstate10D}
\phi^{\vec{h}}(k) \equiv \frac{\overline{\xi^{\vec{h}}} \phi}{\overline{\xi^{\vec{h}}} k^{\vec{h}}}
\end{equation}
for the fermions. The difference in bosonic and fermionic states can be indicated by the integer or half-integer valued eigenvalues $h$. Again, by the usual equivalence theorem all non-linear terms in the transformations can be dropped as they will not contribute to scattering amplitudes after LSZ reduction (they do not have a single particle pole).

Using the analysis of the Clifford algebra given above, it is straightforward to derive the transformation properties of the creation operators up to terms which vanish on-shell,
\begin{equation}\label{eq:trafoprops10D}
\mathfrak{Q} f^{\vec{h}}(k) = \sum_{\vec{h}_1 + \vec{h}_2 = \vec{h}}\left( \overline{ \chi } k^{\vec{h}_2} \right) f^{\vec{h}_1}(k) + \mathcal{O}(k^2=0)
\end{equation}
which corresponds to the transformation derived earlier by representation theory methods. Care should be taken that there is a constraint that all weight vectors should correspond to particles in the multiplet. This is just the usual highest weight state condition. The on-shell supersymmetry algebra can be verified directly. For the transformation of the vector state for instance,
\begin{align}
\mathfrak{Q} e^{\mu}(k, \vec{h}) A_{\mu} =  e^{\mu}(k, \vec{h}) \overline{\chi} \gamma_{\mu} \phi
\end{align}
Now equation \eqref{eq:sumsoverspinors} can be used, which will lead to a sum over $8$ terms. However, all terms which are of the form
\begin{equation}\label{eq:vanfacfermextleg}
\overline{k^{\vec{h}}} \phi
\end{equation}
cannot contribute to scattering amplitudes. This follows because LSZ requires these amplitudes to be connected to a Feynman diagram by a single (renormalized) leg, which will yield a propagator factor
\begin{equation}
\sim \frac{\gamma_{\mu} k^{\mu}}{k^2}
\end{equation}
The denominator will be amputated by LSZ, while the numerator will cause terms of the form \eqref{eq:vanfacfermextleg} to vanish. This leaves four fermion terms which have the form of the fermion creation operators \eqref{eq:extfermstate10D}.

The susy transformation of the fermion state can be worked out similarly. For this it is easiest to expand out $A_{\mu}$ into the complete basis spanned by $k,q$ and the polarization vectors,
\begin{equation}
A_{\mu} = \frac{(k \cdot A) q^{\mu}}{q \cdot k} + \frac{(q \cdot A) k^{\mu}}{q \cdot k} + \sum_{\vec{h}}  \left(e^{-\vec{g}}_{\rho} A^{\rho}\right) e^{\vec{g}}_{\mu}
\end{equation}
It is clear that the first term in this expansion vanishes by the (linearized) gauge symmetry Ward identity, while the second vanishes by the antisymmetry of $\Sigma$. The remaining terms all consist of polarization vectors. Hence the supersymmetry variation of the spinor state reads
\begin{equation}
\delta \phi^{\vec{h}}(k) = -\frac{1}{2} \sum_{\vec{g}} \left( \frac{\overline{\xi^{\vec{h}}}}{\overline{\xi^{\vec{h}}} k^{\vec{h}}} \left(k_{\mu} \gamma^{\mu} e^{\vec{g}}_{\nu} \gamma^{\nu} -  e^{\vec{g}}_{\nu} \gamma^{\nu} k_{\mu} \gamma^{\mu}  \right) \chi \right) \,  e^{-\vec{g}}_{\rho} A^{\rho}
\end{equation}
Up to terms which vanish on-shell, using \eqref{eq:sumsoverspinors} this reads
\begin{equation}
\delta \phi^{\vec{h}}(k) =  \sum_{\vec{g}} \left(\overline{k^{\vec{h}-\vec{g}}} \chi \right) \, e^{-\vec{g}}_{\rho} A^{\rho}
\end{equation}
with the constraint that ${\vec{h}-\vec{g}}$ is a bona-fide spin $\frac{1}{2}$ weight vector. The above can in principle be treated off-shell by appealing to massive spinor helicity. Using the equation \eqref{eq:fourdimlimit} it can be checked that in the $D=4$ limit this expression reproduces the complete usual $\mathcal{N}=4$ multiplet transformations.

\section{Applications to amplitude calculations}
\label{sec:simpleapl}
In this section the simplest examples of all-order amplitudes in higher dimensional Yang-Mills theories are obtained, which generalize known vanishing results in four dimensions. Note that knowledge of the four dimensional massless case can be utilized in several ways. The simplest is the restriction for a higher dimensional gauge theory that when all momenta are restricted to a common $4$ sub-dimensions one should reproduce the known $4$ dimensional results. In general a four dimensional limit like this may be incompatible with the chosen Cartan generators; there is of course a choice generators which is compatible. Amplitudes which involve up to five legs fall into this category as their momenta always span a four dimensional subspace. As discussed above, this choice of Cartan basis is not a gauge transformation in general and this obstruction changes the analysis of the amplitudes in an essential way.

In order to calculate an amplitude through traditional Feynman diagrams one needs the off-shell expressions for the creation and annihilation operators of the states. These were written above in equations \eqref{eq:extvecstate10D} and \eqref{eq:extfermstate10D}, reproduced here for convenience,
\begin{equation}
g^{\vec{h}}(p) \equiv e^{\mu}(k, \vec{h}) A_{\mu} \quad \quad \phi^{\vec{h}}(p) \equiv \frac{\overline{\xi^{\vec{h}}} \phi}{\overline{\xi^{\vec{h}}} k^{\vec{h}}} \nonumber
\end{equation}
for the vector fields $A_{\mu}$ and the fermion fields $\phi$.

\subsection{Vanishing amplitudes at tree level}
Consider the calculation of ordinary Yang-Mills amplitudes at tree level. From \eqref{eq:simplebutpowerful} it is known that there is a class of amplitudes for which the polarization vectors of all the external states are orthogonal. In four dimensions this class includes the helicity equal amplitudes. Calculating this amplitude at tree level therefore reduces to tracking metric factors appearing in the numerator. If a certain diagram has at least one, then the contribution from that diagram to the amplitude vanishes. In Feynman-'t Hooft gauge the numerator of the propagator consists only of a metric. Hence the only source of numerator momentum factors at tree level in this gauge in any minimally coupled, power counting renormalisable field theory are the three point vertices. Simple combinatorics shows that there can be at most $n-2$ momentum factors for an $n$-particle amplitude at tree level. Graphs with this amount of momentum factors consist only of three vertices\footnote{Incidentally, this is also the class of diagrams which is naively at least most in danger of spoiling BCFW recursion.}. Hence every Feynman diagram at tree level in this gauge must contain at least $1$ metric contraction between the polarization vectors.

This argument leads to the natural series of vanishing tree level amplitudes
\begin{equation}\label{eq:vanampstreelevelD10}
\braket{\left(h_1 = \pm 1\right)^{i_1} \ldots \left(h_{\frac{D}{2}} = \pm 1\right)^{\frac{D}{2}}} = 0
\end{equation}
for any numbers of particles $i_1, \ldots i_{\frac{D}{2}}$ and any choice of signs for the different non-trivial weight labels. This is the direct generalization of the vanishing of the helicity equal amplitudes at tree level. Note that the extra vanishing amplitudes obtained here reduce in a four dimensional limit for all particle momenta to amplitudes with for instance chiral scalars, but not their complex conjugate. These amplitudes vanish trivially in the four dimensional case as the vertices always contain the same number of chiral and anti-chiral scalars.

Interestingly, these vanishing tree level amplitudes in higher dimensional Yang-Mills as given above immediately translate into a series of vanishing amplitudes in higher dimensional Einstein gravity through (the field theory limit of) the KLT relations \cite{Kawai:1985xq}. The polarization states here are defined through \eqref{eq:gravpolastates} and their natural generalization. The argument just given can be found for four dimensional Yang-Mills in \cite{Dixon:1996wi}.

The next case in terms of complexity is, in four dimensions, the one helicity unequal amplitude. This corresponds here to adding one gluon to \eqref{eq:vanampstreelevelD10} which is oppositely polarized in the sense of \eqref{eq:simplebutpowerful}. By the same reasoning as above this amplitude consists in Feynman-'t Hooft gauge of a sum over terms proportional to
\begin{equation}
\sim \left(e^{\vec{h}}_i \cdot e^{-\vec{h}}_1 \right) \left(\ldots\right)
\end{equation}
where $1$ labels the special helicity particle. In four dimensions one can employ local gauge invariance to gauge all the like-helicity particles to a gauge in which
\begin{equation}
\left(e^{\pm}_i \cdot e^{\mp}_1 \right) =0 \quad \quad (4\textrm{D})
\end{equation}
Importantly, in this new gauge
\begin{equation}\label{eq:innerprodsaftgaug}
\tilde{e}^{\pm}_i \cdot \tilde{e}^{\pm}_j =0 \quad \quad (4\textrm{D})
\end{equation}
still holds which is manifest in $4D$ spinor helicity notation. Therefore the one helicity unequal amplitude vanishes in four dimensions by, basically, gauge invariance. Although in higher dimensions the `gauge' part of this argument can of course be repeated this does not hold for the second part, displayed in \eqref{eq:innerprodsaftgaug}. One way to make this more precise is to write a generic gauge transformation in terms of pure spinors,
\begin{equation}
e_\mu(\vec{h}) + g k_{\mu}  = \frac{1}{\overline{\xi(\vec{h}_2)} \psi(\vec{h}_2)} \overline{\xi(\vec{h}_2)} \gamma_{\mu} \psi(\vec{h}_1) + g \overline{\psi(\vec{h}_1)} \gamma_{\mu} \psi(\vec{h}_1)
\end{equation}
Where we have chosen one $\vec{h}_1$ for which there is a $\vec{h}_2$ such that $\vec{h} = \vec{h}_1 - \vec{h}_2$. In other words, the gauge transformation can be interpreted to shift
\begin{equation}\label{eq:shiftbygaugetrafo}
\frac{\overline{\xi}(\vec{h}_2)}{\overline{\xi(\vec{h}_2)} \psi(\vec{h}_2)}  \rightarrow  \frac{ \overline{\xi(\vec{h}_2)}}{\overline{\xi(\vec{h}_2)} \psi(\vec{h}_2)} + g \overline{\psi(\vec{h}_1)}
\end{equation}
Now certainly there is a $g$ for which the resulting spinor is annihilated by $\tilde{q}_{\mu} \gamma^{\nu}$ such that the new polarization vector is in $\tilde{q}$ lightcone gauge. However, this gauge transformation does not replace $q$ by $\tilde{q}$ in the polarization vectors since in general there is \emph{no} $g$ such that
\begin{equation}\label{eq:tobesatisfiedforvanishing}
\frac{ \overline{\xi(\vec{h}_2)}}{\overline{\xi(\vec{h}_2)} \psi(\vec{h}_2)} + g \overline{\psi(\vec{h}_1)} \sim \alpha \,  \overline{\psi(\tilde{q}, \vec{h}_2)}
\end{equation}
for some proportionality constant $\alpha$. This follows simply because these are half of the minimal spinor dimension worth of equations, overrestricting $g$ except in four dimensions. Hence the natural generalization of the one helicity unequal amplitude does not vanish in higher dimensions, while it does in four. This can be verified numerically using the techniques of this article. Of course, in the above example for the special choice of gauge of $q_{\textrm{gauge}} = e^{-\vec{h}}_1$ for the helicity equal legs, and any choice but this for the first leg the amplitude does vanish.

Actually, the above equation \eqref{eq:tobesatisfiedforvanishing} could be satisfied if one would be allowed a non-Abelian little group transformation in addition to the gauge transformation. In general these little group transformations are not symmetries of the amplitude however. It does suggest there might be sums over external states to be found which are and it would be interesting indeed to construct these.

\subsection{Vanishing amplitudes from supersymmetric Ward identities}
In a four dimensional supersymmetric theory, the existence of vanishing amplitudes can be extended to any order in perturbation theory by the supersymmetric Ward identities. In \cite{ArkaniHamed:2008yf} this was derived elegantly from the coherent state representation constructed in that paper. The analysis here will be set up completely analogously and is structurally the same. However, just as was found at tree level above there is an essential difference between four and higher dimensions.

To show the vanishing of the analogue of the helicity equal amplitude using coherent states, consider an $n$-point amplitude in coherent state space for which the same top-component given by the weight vector $\vec{h}$ is picked of,
\begin{equation}
A_n\left({k_i, \vec{h}}\right) = \int \prod_{i=1}^{n}d^{\mathcal{D}}\psi_i \tilde{A}_n \left({k_i, \vec{h}, \psi_i}\right)
\end{equation}
where $\mathcal{D}$ is the minimal spinor dimension. The goal now is to use supersymmetry to shift one of the $\psi$, say $\psi_1$ to zero, while only shifting the other $\psi$ variables. As these are all integrated over, this shift will not affect the amplitude and we obtain an integration over $\psi_1$ for which there are no fermionic variables under the integral sign. Hence this amplitude would vanish. The tricky part of this argument, even in four dimensions, is the assumption that there is a supersymmetry transformation which only \emph{shifts} the $\psi$ variables. As displayed in equation \eqref{eq:cohsusytrafo}, a generic supersymmetry variation will lead to phase factors in addition to the shifts. The question is, can one choose a clever variation such that the phase factor
\begin{equation}
\exp\left(\sqrt{2} \sum_{\vec{l} \in L_{\vec{t}}}  \left( \overline{\chi} k^{\vec{h}} \right) \eta_{\vec{l}}  + \sqrt{2} \sum_{\vec{l} \in \overline{L}_{\vec{t}}} \left(\overline{k^{\vec{h}}} \chi \right) \overline{\eta}_{\vec{l}}\right)
\end{equation}
does not contribute?

In four dimensions, one can use chirality to split the supersymmetry transformations into chiral parts. Suppose only chiral or anti-chiral variables are used to describe the multiplet (which corresponds to choosing either the $+$ or $-$ helicity gluon state as the top state). It is easy to see that when one applies a chiral or anti-chiral transformation the phase factor is automatically absent. Although this seems natural, this supersymmetry transformation is inherently complex and use of this is in fact only allowed if $U(1)_R$ is unbroken. One way to test if this symmetry is unbroken is to calculate fermion helicity violating amplitudes, which perturbatively indeed turn out to vanish in all known cases. Hence in four dimensions, using chiral supersymmetry transformations there are twice the number of supersymmetry generators as degrees of freedom of the susy transformations which do not lead to phase factors in the transformations of the above coherent state amplitude. For $\mathcal{N}=4$ for instance, one can with these transformations generically satisfy $8$ conditions on the coherent state variables.

In higher dimensions, the situation is more restricted. Here one cannot substitute chirality for the relevant quantum numbers as the little group is non-Abelian. By \eqref{eq:propfactspinprod} it is easy to pick a supersymmetry variation which leaves the conjugate weights invariant: choose the variation
\begin{equation}\label{eq:nophasefac}
\chi \in \textrm{span}\{\xi(h\in L_{\vec{t}})\} \quad \quad \overline{\chi} \in \textrm{span}\{\xi(h\in \overline{L}_{\vec{t}})\}
\end{equation}
This supersymmetry variation will not lead to phase factors. The question is if there are other variations for which the phase factor vanishes for all legs. By expanding $\chi$ in a basis of spinors spanned by $\xi$ and $\psi(k_1)$ and trying to solve the resulting equations particle by additional particle it follows that there is no such $\phi$ except for \eqref{eq:nophasefac}: at some point there are more equations than degrees of freedom.

Hence in higher dimensions than four there is only the same number of degrees of freedom in supersymmetry transformations as fermionic coordinates. Special choices of complex structure can in principle alleviate this situation. The upshot of the whole analysis is that it is easy to show that the amplitude with all weight labels equal vanishes by shifting the coherent state variables of one particle to zero. However, supersymmetry transformations in higher dimensions are unable to resolve the next case which in four dimensions corresponds to the one helicity unequal amplitude. This confirms what was found at tree level above.

The same argument can be applied to a wider class of amplitudes: those formed by the top components of a coherent state representations and those particles which have one coherent state parameter label in common with this top state. Interestingly, this does not cover the class of amplitudes in \eqref{eq:vanampstreelevelD10}, but only those which have same sign vector. If the amplitude involves fermions, the fermion Dynkin labels should have a Dynkin label with the same sign. In four dimensions these amplitudes violate $R$-symmetry. Furthermore the argument holds for the top component of a coherent state representation in any supersymmetric higher dimensional field theory. Since the supersymmetric Ward identities do not depend on any coupling constant, this proves vanishing amplitudes to all orders in perturbation theory. As a simple example, the argument applies to $D=10$ super Yang-Mills and $D=11$ super-gravity.

As pointed out in the four dimensional context \cite{Boels:2008fc} (motivated by results obtained in a more round-about way in \cite{Stieberger:2007jv} and spelling out a line of thought also used in \cite{Berkovits:2008ic}), one non-trivial coupling constant one can consider are the string scale $\alpha'$ and the string coupling constant $g_s$. The Ward identities derived in this article therefore apply to full string amplitudes in a flat $10$ dimensional background (or to amplitudes on the world-volume of a stack of flat branes). In particular, the vanishing amplitudes with only gluon legs discussed above also apply in the string theory, greatly expanding the list of known amplitudes for the superstring to all multiplicities. This can be checked explicitly for the six point disc amplitude in \cite{Oprisa:2005wu}, as every kinematic structure appearing there contains at least one metric contraction. By the KLT relations, this means that the closed string sphere amplitudes for gravitons with helicities specified as above also vanish. This argument can be extended to amplitudes which involve both open and closed string modes by using \cite{Stieberger:2009hq}. For applications which involve closed strings one has to use \eqref{eq:gravpolastates} to translate between spin $1$ and spin $2$ particles.

\section{Discussion and conclusions}
Despite massive progress in the analytic understanding of scattering amplitudes of theories with massless vector particles in four dimensions, not nearly so much has been done for their massive and higher dimensional counterparts. As this is a rich area of physics with applications ranging from phenomenology to more formal theory, there is a definite motivation to do so. In this article we have made a first step toward the extension to higher dimensions by a complete and, crucially, covariant discussion of the Poincar\'e algebra quantum numbers of the legs in any scattering amplitude. Both vector as well as spinor-type representations have been discussed resulting in a natural dictionary between the two. This dictionary constitutes a complete spinor helicity method for higher dimensional field theory. The mapping is not one-to-one however: in dimensions higher than four there are for a given polarization multiple ways to express it in terms of spinors. This redundancy might be indicative of a symmetry which can be further exploited. A deeper, more geometrical understanding of the construction for instance in the direction of pure spinor spaces could help here.

It is intriguing to see pure spinors arise naturally in the analysis. The space of pure spinors is the space of all inequivalent complex structures which can be put on a $D$-dimensional real space. In four dimensions, this is twistor space and it is easy to speculate that the space of pure spinors can play a similar role in higher dimensions as twistor space in four dimensions. See e.g. \cite{Berkovits:2004bw} and references therein for some results along these lines. Here we have made a further step in this direction by elucidating the role pure spinors can play in defining on-shell polarization states. How far the analogy to recent developments in (ambi-)twistor space in four dimensions as in for instance \cite{ArkaniHamed:2009si, Mason:2009sa} can really be pushed is an outstanding and highly interesting question. Finding an analog of the half-Fourier transform would be a good starting point here.

Apart from direct applications to field theory, another research direction to be explored is string theory. Pure spinors entered the vocabulary of theoretical physics following Berkovits's pure spinor based approach to the covariant quantization of the superstring \cite{Berkovits:2000fe}. The interplay between the string theory and the pure spinor based representation theory described in this paper should be a fertile ground to explore further the application of recently discovered field theory techniques to the string and of string derived ideas to the field theory. The fairly simple expressions for the tree level four point superstring amplitudes in a flat background summarized in \cite{Mafra:2009wq} might be a good starting point for this.

\begin{figure}[t]
  \begin{center}
  \includegraphics[width=0.5\textwidth]{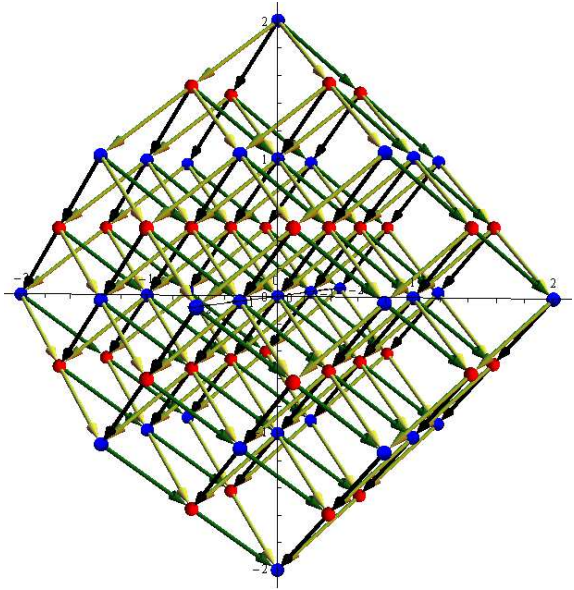}
  \caption{the $\mathcal{N}=1$, $D=11$ multiplet in projected weight space, ignoring particle multiplicities}
  \label{fig:n8multiplet}
  \end{center}
\end{figure}

The methods discussed in particular allow a completely on-shell derivation of explicit supersymmetry Ward identities for scattering amplitudes with in principle arbitrary field content, in arbitrary (flat) dimensions for any supersymmetric field theory without central charges. The extension to theories with central charges is in principle straightforward. It would be interesting to see if these techniques can for instance be applied to the polarization sums which appear in computing higher loop scattering amplitudes in maximally supersymmetric field theories in four dimensions. There has been substantial work for the four dimensional sums, see e.g. \cite{Bern:2009xq} and references therein. One of the applications of higher dimensional techniques in this arena are direct high loop amplitude calculations for four dimensional $\mathcal{N}=8$ supergravity, whose rather involved multiplet structure is sketched in figure \ref{fig:n8multiplet}.

The applications of the developed technology to concrete amplitudes presented above should be only a tip of an iceberg. This tip consists of vanishing amplitudes to all multiplicities, all orders. In the four dimensional massless case this result is related to the existence of an integrable sub-sector in the theory: self-dual Yang-Mills theory. A natural generalization\footnote{This was pointed out to me by David Skinner.} would be to pose that higher dimensional theories also have a similar integrable sub-structure. This would raise the possibility of a CSW-style \cite{Cachazo:2004kj} perturbation theory around a self-dual sector. However, there is an important issue here. One way to interpret the non-vanishing of the one-helicity-unequal amplitudes in four dimensions was given in \cite{Mason:2008jy}: these amplitudes measure linear perturbations of a self-dual (BPS) background. Since these vanish in four dimensions one can, in this number of dimensions, consistently truncate both gravity and Yang-Mills theory to a self-dual theory. MHV amplitudes can then be interpreted as quadratic fluctuations in the stable self-dual background. By a similar reasoning there therefore appears to be no consistent similar truncation to a self-dual sector in higher dimensions as the amplitudes associated to linear perturbations do not vanish. There might be more involved choices of external states which would reveal an integrable structure.

Obtaining non-vanishing amplitudes in general remains of course a prime objective of further research. The analogue of the one helicity unequal amplitude in four dimensions is the natural candidate to study first, as simple kinematic reasoning show this class of amplitudes closes on itself under on-shell recursion \cite{Britto:2004ap, Britto:2005fq}. As might be expected on-shell recursion takes a very natural form in terms of pure spinors. This is work in progress.

\acknowledgments
It is a pleasure to thank Christian Schwinn, Zvi Bern, Emil Bjerrum-Bohr, Niels Obers and David Skinner for discussions and/or comments. The research reported in this paper is supported by a Marie Curie European Reintegration Grant within the 7th European Community Framework Programme.

\bigskip

\appendix

\section{Spinors sums and phase conventions}
\label{app:eigenspinornorm}
In this appendix it is shown how to obtain the normalization of the spinors leading to equation \eqref{eq:sumsoverspinors} from the representation theory of the fermionic harmonic oscillator.

Fixing the scaling ambiguity of equation \eqref{eq:scalingambispinors} amounts to a choice of conventions for the phases of the spinor wave functions. A convenient and consistent, but by no means unique set can be fixed by a choice of highest weight state $\xi_{\textrm{top}}$,
\begin{equation}\label{eq:highweightchoice}
\begin{array}{cl}\gamma_i^{+} \xi_{\textrm{top}}\left(\vec{h}_{\textrm{top}} \right) & = 0  \\
\gamma_{\mu} q^{\mu} \xi_{\textrm{top}}\left(\vec{h}_{\textrm{top}}\right) & = 0 \ .
\end{array}
\end{equation}
with $\vec{h}_{\textrm{top}} = \left(\frac{1}{2},\ldots,\frac{1}{2}\right)$. The phases of the other states in the theory can then be \emph{defined} by application of successive lowering generators ordered to index, e.g.
\begin{align}
\xi\left(-\frac{1}{2},\frac{1}{2}, \ldots,\frac{1}{2} \right) & \equiv \gamma_1^{-} \xi_{\textrm{top}}\left(\vec{h}_{\textrm{top}}\right) \\
\xi\left(\frac{1}{2},-\frac{1}{2}, \ldots,\frac{1}{2} \right) & \equiv \gamma_2^{-} \xi_{\textrm{top}}\left(\vec{h}_{\textrm{top}} \right) \\
\xi\left(-\frac{1}{2},-\frac{1}{2}, \ldots,\frac{1}{2} \right) & \equiv \gamma_1^{-} \gamma_2^{-} \xi_{\textrm{top}}\left(\vec{h}_{\textrm{top}}\right)
\end{align}
The action of the other generators of the algebra on the states follow by the algebra and the annihilation conditions \eqref{eq:highweightchoice}. For complex momenta, one can pick a conjugate highest weight (now taken to be independent)
\begin{equation}\label{eq:highweightchoiceII}
\begin{array}{cl}  \overline{\xi_{\textrm{top}}\left(\frac{1}{2},\ldots,\frac{1}{2} \right)} \gamma_i^{-} & = 0  \\
 \overline{\xi_{\textrm{top}}\left(\frac{1}{2},\ldots,\frac{1}{2} \right) } \gamma_{\mu} q^{\mu} & = 0 \ .
\end{array}
\end{equation}
and proceed analogously. With the chosen convention conventions it is easy to write explicit forms of all the operators in the theory in terms of the inner product \eqref{eq:propfactspinprod},
\begin{align}
\gamma_0^{-} & = \sum_{\vec{h}} \frac{\psi^{\vec{h}} \overline{\psi^{\vec{h}}}}{ \overline{\psi^{\vec{h}}} \xi^{\vec{h}}} \\
\gamma_0^{+} & = \sum_{\vec{h}} \frac{\xi^{\vec{h}} \overline{\xi^{\vec{h}}}}{ \overline{\xi^{\vec{h}}} \psi^{\vec{h}}}
\end{align}
and
\begin{align}\label{eq:polvecsexp}
e_{\mu}\left(\vec{f}_i = \pm \delta_i^j\right) & = \sum_{\vec{h} | h^j = \mp \frac{1}{2}} \left(-1\right)^{\sum_{k=1}^i h_j} \left( \frac{\psi(\vec{h} + \vec{f}) \overline{\xi(\vec{h})}}{\overline{\xi(\vec{h})} \psi(\vec{h})} - \frac{\xi(\vec{h} + \vec{f}) \overline{\psi(\vec{h})}}{\overline{\psi(\vec{h} + \vec{f})} \xi(\vec{h} + \vec{f})} \right)
\end{align}
These equations express the fact all the operators in \eqref{eq:completegammabasis} are nilpotent and therefore can only have non-trivial eigenvectors with vanishing eigenvalue. The complete non-trivial eigenvector space is spanned by the solutions to the Dirac equation which are labeled by the weight vectors. The above equations simply imply a normalization convention for the fermions.

For real momenta the highest weight state phase convention leads to all inner products being equivalent, since for instance
\begin{equation}
\overline{\xi\left(\vec{h}\right)} \psi\left(\vec{h}\right) = \overline{\xi\left(\vec{h}\right)} \gamma_i^{\pm} \gamma^{\mp}_i \psi\left(\vec{h}\right)  =  \overline{\xi\left(\vec{h}+ \vec{h}_i\right)} \psi\left(\vec{h} + \vec{h}_i \right) \quad \textrm{(real momenta)}
\end{equation}
the first equality sign follows from the algebra. The same is not true for complex momenta, but a modified statement tracking the phases can be found.

With the above consistent set of phase conventions, others (such as the one used in the main text) may be constructed. First consider redefining all states in the theory as
\begin{eqnarray}
\psi(\vec{h}) \rightarrow \alpha_{\vec{h}} \psi(\vec{h}) & \quad \quad &  \xi(\vec{h}) \rightarrow  \beta_{\vec{h}} \xi(\vec{h})\\
\overline{\psi(\vec{h})} \rightarrow  \overline{\alpha}_{\vec{h}} \overline{\psi(\vec{h})} & \quad \quad & \overline{\xi(\vec{h})} \rightarrow \overline{\beta}_{\vec{h}} \overline{\xi(\vec{h})}
\end{eqnarray}
If the scale factors are chosen to obey
\begin{align}
\alpha_{\vec{h}} \overline{\alpha}_{\vec{h}} & = \overline{\psi(\vec{h})} \xi(\vec{h}) \\
\beta_{\vec{h}} \overline{\beta_{\vec{h}}} & = \frac{2 q \cdot k}{\overline{\xi(\vec{h})} \psi(\vec{h})}
\end{align}
or equivalently in terms of the new spinors
\begin{align}\label{eq:normspinspec}
\frac{\alpha_{\vec{h}}}{\beta_{\vec{h}}} & = \overline{\psi'(\vec{h})} \xi'(\vec{h}) \\
\frac{\beta_{\vec{h}}}{\alpha_{\vec{h}}} & = \frac{2 q \cdot k}{\overline{\xi'(\vec{h})} \psi'(\vec{h})}
\end{align}
then
\begin{equation}\label{eq:normconds}
\left(\begin{array}{rc}
k_{\mu} \gamma^{\mu} & = \sum_{h} \psi^{\vec{h}} \overline{\psi^{\vec{h}}} \\
q_{\mu} \gamma^{\mu} & = \sum_{h} \xi^{\vec{h}} \overline{\xi^{\vec{h}}}
\end{array}\right.
\end{equation}
where we have dropped the prime's for notational convenience. Next up are the slashed polarization vectors. These will acquire factors of $\alpha(\vec{h})$ and $\beta(\vec{h})$ which can be used to set,
\begin{equation}
\left(-1\right)^{\sum_{k=1}^i h_j} \frac{\alpha(\vec{h}+ \vec{f})}{\alpha(\vec{h})}= 1
\end{equation}
so that we arrive at
\begin{align}
e_{\mu}\left(\vec{f}_i = \pm \delta_i^j\right) & = \sum_{\vec{h} | h^j = \mp \frac{1}{2}}   \left( \frac{\psi(\vec{h} + \vec{f}) \overline{\xi(\vec{h})}}{\overline{\xi(\vec{h})} \psi(\vec{h})} - \frac{\xi(\vec{h} + \vec{f}) \overline{\psi(\vec{h})}}{\overline{\psi(\vec{h} + \vec{f})} \xi(\vec{h} + \vec{f})} \right)
\end{align}

For complex momenta $\alpha_{\vec{h}}$ and $\overline{\alpha_{\vec{h}}}$ are uncorrelated, and we can choose all $\alpha(\vec{h})$ and $\beta(\vec{h})$ to be proportional to the signs in \eqref{eq:polvecsexp}, while $\overline{\alpha}(\vec{h})$ and $\overline{\beta}(\vec{h})$ solve the above conditions. The same choice is also possible for real momenta, when the above conditions are degenerate. Therefore, in the set of phase conventions given by \eqref{eq:normconds} and \eqref{eq:polvecsexp} there is now an overall phase ambiguity left for both the $\psi$ type spinors if complex momenta are used. In addition, the normalization of the $\xi$ spinors is undetermined in the sense that any transformation
\begin{equation}
\xi(\vec{h}) \rightarrow  \beta_{\vec{h}} \xi(\vec{h}) \quad \quad \overline{\xi(\vec{h})} \rightarrow \overline{\beta}_{\vec{h}} \overline{\xi(\vec{h})}
\end{equation}
for which
\begin{equation}
\beta_{\vec{h}} \overline{\beta}_{\vec{h}} = 1
\end{equation}
does not change the form of the generators. It will affect the form of the spinor inner product by a phase.

Note that with the conventions chosen
\begin{equation}
\left(\overline{\psi(\vec{h})} \xi(\vec{h}) \right) \, \left( \overline{\xi(\vec{h})} \psi(\vec{h}) \right)= 2 q \cdot k
\end{equation}
holds for the re-scaled spinors by equation \eqref{eq:normspinspec}, for every weight vector $\vec{h}$.

\section{Compendium of spinor helicity formulas}
This section contains as a service to the reader a quick overview over all the formulas which together form a pure spinor helicity method. Explanations in the main text.

\subsection*{Initial choices}
Choose number of dimensions $D$ and a frame
\begin{equation}
q, \hat{q}, n_i
\end{equation}
for $i=1,\ldots,D-2$ such that the only non-trivial inner products between these are
\begin{equation}
\hat{q} \cdot q = n_i^2 = 1
\end{equation}
The vector $q$ is chosen to have non-vanishing inner products with all the momenta in the problem. Choose furthermore a complex structure as
\begin{align}
z_i & = \frac{n_{2i-1} + \ii n_{2i}}{\sqrt{2}} \\
\bar{z}_i & = \frac{n_{2i-1} - \ii n_{2i}}{\sqrt{2}}
\end{align}
with $i=1 \ldots \frac{\left(D-2\right)}{2}$.

\subsection*{Vector polarization states}
\subsubsection*{massless case}
First suppose momentum $k$ is light-like, $k^2=0$. Construct new vectors which span the space orthogonal to both $q$ and $k$ as
\begin{equation}
\tilde{n}^i  = n^i - q \frac{n^i \cdot k }{q \cdot k}
\end{equation}
Using the complex structure this yields
\begin{align}
m^i & = \frac{\tilde{n}^{2i} + \ii \tilde{n}^{2i +1}}{\sqrt{2}} \\
\bar{m}^i & = \frac{\tilde{n}^{2i} - \ii \tilde{n}^{2i +1}}{\sqrt{2}}
\end{align}
as $q$-lightcone gauge polarization vectors which have definite, simultaneous eigenvalues under
\begin{equation}
R^j_q = \ii m_j^{\mu} \bar{m}_j^{\nu} \Sigma^V_{\mu \nu}
\end{equation}
as
\begin{align}
R_q^j \, m^i & = \delta_{ij} m^i \equiv  h^j m^i \\
R_q^j \, \bar{m}^i  & = - \delta_{ij} \bar{m}^i  \equiv h^j \bar{m}^i
\end{align}
Here $\Sigma^V_{\mu \nu}$ is the rotation matrix in the vector representation. The eigenvalues under $R^j_q$ form a weight vector $\vec{h}$. The corresponding polarization vectors obtained above will be denoted by
\begin{equation}
e_{\mu}(k, \vec{h})
\end{equation}
The set
\begin{equation}
\{q\, , k\, , e_{\mu}(k, \vec{h}) \}
\end{equation}
with the $D-2$ states labeled by $\vec{h}$ forms a complete basis of the vector space.

\subsubsection*{massive case}
Given a momentum $k$, construct massless momentum $k^{\flat}$ as
\begin{equation}
k_{\mu} = k^{\flat}_{\mu} + \frac{k^2}{2 q\cdot k } q_{\mu}
\end{equation}
and proceed as above. The extra polarization in this case can be given as
\begin{equation}
e^0_{\mu} = k^{\flat}_{\mu} - \frac{k^2}{2 q\cdot k } q_{\mu}
\end{equation}
The resulting polarization vectors are in unitary gauge.

\subsection*{Spinor polarization states}
\subsubsection*{massless}
There are spinors $\psi(\vec{h})$ and $\xi(\vec{h})$ which obey
\begin{equation}
\begin{array}{cl}\gamma_i^{2 h^i} \psi(\vec{h}) & = 0  \quad \quad \textrm{no sum}\ , \\
\gamma_{\mu} k^{\mu} \psi(\vec{h}) & = 0
\end{array}
\end{equation}
as well as
\begin{equation}
\begin{array}{cl}\gamma_i^{2 h^i} \xi( \vec{h}) & = 0  \quad \quad \textrm{no sum}  \\
\gamma_{\mu} q^{\mu} \xi(\vec{h}) & = 0 \ .
\end{array}
\end{equation}
and are normalized such that
\begin{equation}
\begin{array}{ccc}
\gamma^{\mu} k_{\mu}  & = & \sum_{\vec{h}} \psi^{\vec{h}} \overline{\psi^{\vec{h}}} \\
\gamma^{\mu} q_{\mu}  & = & \sum_{\vec{h}} \xi^{\vec{h}} \overline{\xi^{\vec{h}}} \\
\gamma^{\mu} e_{\mu}\left(\vec{f}_i = \pm \delta_i^j\right) & = & \frac{\sqrt{2}}{\ii} \sum_{\vec{h} | h^j = \mp \frac{1}{2}} \left( \frac{\psi(\vec{h} + \vec{f}) \overline{\xi(\vec{h})}}{\overline{\xi(\vec{h})} \psi(\vec{h})} - \frac{\xi(\vec{h} + \vec{f}) \overline{\psi(\vec{h})}}{\overline{\psi(\vec{h} + \vec{f})} \xi(\vec{h} + \vec{f})} \right)
\end{array}
\end{equation}
with the bar denoting the usual spinor conjugation for real momenta. For complex momenta the conjugate spinors should be treated as independent. These spinors form a complete basis of the spinor space. In the massless case there can be Majorana or Weyl reduction conditions.

\subsubsection*{massive}
With the same decomposition as for massive vectors,
\begin{equation}
\psi(k, \vec{h}) = \psi(k^{\flat}, \vec{h}) + \frac{m}{\overline{\psi(k^{\flat},\vec{h})} \xi(\vec{h})} \xi(\vec{h})
\end{equation}
is a solution to the massive Dirac equation with the indicated eigenvalues. The massless spinors on the right hand side were obtained above. Note a Weyl condition is impossible in the massive case.

\subsection*{vectors in terms of spinors}
Given a vector weight $\vec{h}$, there are $2^{\frac{D}{2}-1}$ different representations in terms of the above defined spinors,
\begin{equation}
e^{\mu}(\vec{h}_1 - \vec{h}_2) = \frac{\ii}{\sqrt{2}} \frac{\overline{\xi(\vec{h}_1)} \gamma^{\mu} \psi(k,\vec{h}_2)}{\overline{\xi(\vec{h}_1)} \psi(k, \vec{h}_1)}
\end{equation}
For every possible Weyl and/or Majorana condition, the above counting gets reduced by a factor of $2$ each. In addition,
\begin{align}
k_{\mu} & = \frac{1}{2} \overline{\psi(\vec{h})} \gamma_{\mu} \psi(\vec{h}) \\
q_{\mu} & = \frac{1}{2} \overline{\xi(\vec{h})} \gamma_{\mu} \xi(\vec{h})
\end{align}
for any weight vector $\vec{h}$ with a similar counting of the possibilities.

\bibliographystyle{JHEP}

\bibliography{susbiblio}

\end{document}